\documentclass[a4paper]{article}

\usepackage[english]{babel}
\usepackage[utf8]{inputenc}
\usepackage{amsmath}
\usepackage{graphicx}
\usepackage[colorinlistoftodos]{todonotes}
\usepackage[]{mcode}
\usepackage[margin=0.8in]{geometry}
\usepackage{mathtools}
\usepackage{float}
\usepackage{tikz}
\usepackage{subfigure}
\usepackage{amssymb}
\usepackage{hyperref}
\usepackage{cleveref}
\usepackage{slashed}
\usepackage{cancel}
\usepackage{blindtext}
\usepackage{cite}
\usepackage{multicol, caption}
\usepackage{dsfont} 

\newcommand{\andspace}{\hspace{3mm} \text{and} \hspace{3mm}}

\newcommand{\U}{\mathcal{U}}
\newcommand{\vac}{|\Omega\rangle}

\newcommand{\V}{\mathcal{V}}

\newcommand{\A}{\mathcal{A}}
\newcommand{\M}{\mathcal{M}}
\newcommand{\R}{\mathcal{R}}

\newcommand{\Prop}{\mathcal{P}}
\newcommand{\Sing}{\mathcal{S}}
\newcommand{\D}{\mathcal{D}}
\newcommand{\tr}{\text{Tr}}
\newcommand{\E}{\mathcal{E}}

\newcommand{\ortho}[1]{\textcolor{gray}{+ \bigg(\perp #1\bigg)}}

\newenvironment{Figure}
  {\par\medskip\noindent\minipage{\linewidth}}
  {\endminipage\par\medskip}

\title{
\textbf{Quantum computation of partonic Drell-Yan scattering cross sections and interference effects}
}
\author
{
Erik Bashore$^{1, 2}$\footnote{\texttt{Erik.Bashore@uib.no}},
~Stefano Moretti$^{1,3}$\footnote{\texttt{stefano.moretti@cern.ch}},
~Timea Vitos$^{1,4}$\footnote{\texttt{timea.vitos@physics.uu.se}}
\vspace{5mm}
\\
{\small\it $^{1}$ Department of Physics and Astronomy, Uppsala University,} 
\\%
{\small\it Box  516,  751 20, Uppsala, Sweden  }\\
{\small \it $^{2}$ Department of Informatics, University of Bergen} \\
{\small \it HIB - Thormøhlens gate 55, Norway}
\\%
{\small\it $^{3}$ School of Physics and Astronomy, University of Southampton,} 
\\%
{\small\it Highfield, Southampton SO17 1BJ, United Kingdom  }\\
{\small\it $^{4}$ Institute for Theoretical Physics, ELTE  E\"otv\"os Lor\'and  University,} 
\\%
{\small\it P\'azm\'any  P\'eter  s\'et\'any  1/A,  H-1117  Budapest,  Hungary}
}

\date{}

\begin{document}
\maketitle
\vspace{-5mm}
\begin{abstract}
    \normalsize We probe the possibilities of efficiently constructing simple Feynman diagrams into quantum devices. More precisely, we study Drell-Yan lepton pair creation at the partonic level of the form $q\bar{q} \to \gamma/ Z \to \ell^-\ell^+$. We develop quantum gates that build up the relevant diagrams using simple Feynman rules, such as vertex and propagator gates $\mathcal{V}$ and $\mathcal{P}$. We show how the quantum circuit may compute simultaneous amplitudes in the phase space and how to reach the full integrated cross-section from the outputs. In addition to this we also show how the circuit is able to simultaneously isolate the interference effects of the contributing diagrams by a simple basis rotation. The circuit design is made to be general, and thus this work constitutes a step towards the implementation of arbitrary scattering process computations and efficient interference analyses. 
\end{abstract}

\begin{center}
\end{center}





\color{black} 

\begin{multicols}{2}

\begin{center}
    \section{Introduction}
\end{center}
The use of Quantum Computing (QC) in High Energy Physics (HEP) has grown from a speculative
idea into an active research direction, primarily driven by the increasing computational demands of
modern collider experiments. As datasets expand and theoretical predictions require ever
greater precision, it becomes natural to explore whether quantum devices can offer advantages
over classical algorithms for selected tasks. Several studies have already demonstrated that QC
can be applied to problems in lattice simulation \cite{Klco:2021lap,Funcke:2023jbq,Yamamoto:2022jnn,Meurice:2020pxc},  loop computations of Feynman graphs \cite{deLejarza:2024pgk,Ramirez-Uribe:2021ubp,Clemente:2022nll},  effective field theories \cite{Bauer:2021gup},  Parton Distribution Functions (PDFs) \cite{Perez-Salinas:2020nem,Li:2024zsw},  parton shower algorithms \cite{Bepari:2020xqi,Bepari:2021kwv}, integration of amplitudes \cite{Agliardi:2022ghn,Williams:2025hza} and event generators in general \cite{Gustafson:2022dsq,Bravo-Prieto:2021ehz,Kiss:2022pjw}, all suggesting
that quantum methods may eventually complement conventional computational tools. Indeed, a transversal activity in the form of the QC4HEP working group \cite{DiMeglio:2023nsa} is gaining momentum within the HEP community at large.

One of the areas where computational complexity becomes particularly visible is the
evaluation of scattering amplitudes in Quantum Chromodynamics (QCD), which involve large color spaces and numerous helicity configurations, so that 
their computational cost grows rapidly with the number of external particles. Even for 
tree-level processes the factorial growth of graph permutations and color structures can dominate
the runtime of event generators. Recent quantum algorithms for gluon amplitudes have shown
that both color factors and helicity structures can be encoded in quantum circuits, providing
a proof-of-concept that gauge-theory ingredients can be represented within a unitary 
framework, see Refs. \cite{Bepari:2020xqi,Bauer:2022hpo,DiMeglio:2023nsa,Chawdhry:2023jks,Bashore:2025uwb}.\\ \indent
Against this broader backdrop, the Drell-Yan (DY) process offers an ideal testing ground for quantum approaches tackling other aspects of scattering amplitudes (other works that are in a similar vein, although with different applications, are for instance found in Refs. \cite{Agliardi:2022, Williams:2025, varona:2024}). In fact, although DY is structurally simple, being a two-body scattering process mediated by Electro-Weak (EW) bosons through two Feynman diagrams in the Standard Model (SM), with a trivial color structure, it contains features that are essential for
realistic collider phenomenology. These include the interplay between a photon and the $Z$ boson through the appearance of interference terms
that can enhance or suppress the cross-section depending on the kinematic regime. Such
interference patterns are central to precision studies as well as searches for physics Beyond the
SM (BSM) \cite{Feuerstake:2025, Accomando:2013, mistlberger:2025}, thereby making DY a meaningful benchmark for quantum simulations.
\\ \indent
The motivation for focusing on DY is therefore twofold. First, its modest diagrammatic
structure allows one to translate Feynman rules directly into quantum gates, enabling a
transparent mapping between field-theoretic ingredients and circuit components. Second, the
process is sufficiently rich to test key capabilities of quantum devices: implementing Dirac
matrices, propagators and couplings, preparing spinor states, isolating interference through
basis rotations as well as evaluating many phase-space points simultaneously by exploiting quantum superposition. The ultimate goal is to assess whether a quantum circuit can simultaneously estimate the full cross-section as well as the isolated interference effects more efficiently and/or accurately than standard deterministic approaches exploiting classic hardware.
\\ \indent
To this end, in this work, we construct a quantum circuit that performs the partonic DY computation
$q\bar{q} \to \gamma/Z \to \ell^+\ell^-$ (i.e., no PDF is considered). The circuit is designed to be general, meaning that its
building blocks (i.e., vertex, propagator and spinor gates as well as index-handling operations) 
can be reused for more complex scattering processes. By embedding the amplitude in a
quantum state and extracting both the full result and the isolated interference contribution, we
aim to assess how quantum devices might eventually contribute to further collider phenomenology in the direction of better modeling sub-leading terms (as interferences normally are) in an amplitude squared calculation.
The DY process thus serves as a controlled environment in which to evaluate the strengths and
limitations of quantum-circuit-based amplitude calculations.

The structure of the paper is as follows. In section 2, we describe the quantum gates and present how the computation of the DY process is performed. The results follow in section 3 and we present our summary and conclusions in section 4. 


\vspace{3mm}
\begin{centering}
\section{The quantum circuit}
\end{centering}
To compute the DY scattering process we wish to implement the corresponding Feynman rules for the contributing diagrams in terms of quantum gates acting on different quantum registers. In this section we list the different gates that we have constructed for this task and give their corresponding action, however, we leave the more in-depth details to Appendix A. Recall that the Feynman rules of the matrix elements $\M_\gamma$ and $\M_Z$ contain three primary pieces: the in/out-going spinors, the vertex factors and the internal bosonic propagator. Each of these get their own dedicated quantum gate. The circuit is realized on a set of qubit registers $\mathcal{R} = \big\{v_1, v_2, i_1, i_2, \U, a_1, a_2, a_3, p\big\}$ where each register refers to $v=$ vertex, $i=$ index, $\U=$ unitarity, $a=$ ancilla and $p=$ particle respectively. The details of these registers, including number of acting qubits, will be clear in time. Throughout this paper we will refer to the full set of registers with $p$ removed, i.e., $\mathcal{R}\setminus p$, as work and also denote $\vac_r \equiv |0\rangle^{\otimes n_r}$ to be the vacuum state of any register $r$ containing $n_r$ qubits.

\end{multicols}

\begin{figure}[h]
\centering
\begin{minipage}[t]{0.32\linewidth}
    \centering
    \subfigure[Spinor gate]{%
        \includegraphics[width=0.55\linewidth]{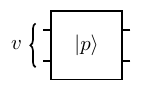}
        \label{fig:spinorgate}
    }

    \vspace{0.6em}

    \subfigure[Barred spinor gate]{%
        \includegraphics[width=0.55\linewidth]{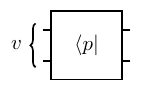}
        \label{fig:antispinorgate}
    }
\end{minipage}
\hspace{-0.11\linewidth} 
\begin{minipage}[t]{0.62\linewidth}
    \centering
    \subfigure[Vertex gate]{%
        \includegraphics[width=0.4\linewidth]{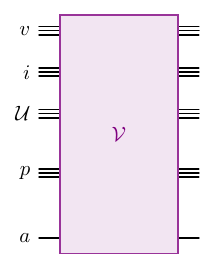}
        \label{fig:vertexgate}
    }
    \subfigure[Propagator gate]{%
        \includegraphics[width=0.4\linewidth]{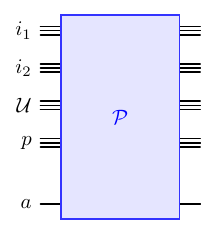}
        \label{fig:propgate}
    }
\end{minipage}

\caption{Circuit diagram representations of the primary quantum gates.}
\label{fig:quantumgates}
\end{figure}

\begin{multicols}{2}
The first gate we introduce is the spinor gate $U_p$ and its barred counterpart $\bar{U}_p$. This gate takes an arbitrary four-component spinor $u_s(\boldsymbol{p})$, with three-momentum $\boldsymbol{p}$ and helicity $s$, and produces a two-qubit quantum state which has the normalized components $\hat{u}_j \equiv u_j/|u_s(\boldsymbol{p})|$ encoded into the probability amplitudes. This gate operates on one of the vertex registers $v_i$ and the action on the vacuum can be written as 
\begin{equation}
U_p\vac_{v_i} = \sum_{j = 0}^3\hat{u}_j|j\rangle_{v_i} \equiv |p\rangle_{v_i}
    \label{spinorgateaction}
\end{equation}
where the new state is spanned by the two-qubit computational basis states $|j\rangle$. Similarly, the barred spinor gate $\bar{U}_p$ produces a state with the normalized barred spinor components onto the dual vector space. Since a barred spinor is found by $\bar{u}_s(\boldsymbol{p}) = u_s^\dagger(\boldsymbol{p})\beta$ the gate can be constructed as $\bar{U}_p = U^\dagger_p \beta$ using a two-qubit gate representation of the $\beta$ matrix. The circuit diagram representations of both the gates can be seen in \crefrange{fig:spinorgate}{fig:antispinorgate} and their explicit construction can be found in Appendix A.

Next we introduce the vertex gate $\V$. This gate is responsible for implementing the corresponding vertex factor of a $Vf\bar{f}$ interaction with internal boson $V$ and in/out-going fermion/anti-fermion pair $f\bar{f}$.
The vertex factor in the EW theory for this interaction has the specific structure 

\begin{equation}
    \V_\mu^{(V)} = C_\V^{(V)} \gamma_\mu + C_\A^{(V)} \gamma_\mu \gamma_5
    \label{vertexfactor}
\end{equation}
with vector and axial-vector couplings $C_\V^{(V)}$ and $C_\A^{(V)}$ respectively. The vertex gate acts upon the registers $\{v, i, \U, a, p\}$ by placing the Dirac bilinears $\gamma_\mu$ and $\gamma_\mu\gamma_5$ onto the vertex register with the spacetime index $\mu$ controlled by the index register. The values of the couplings are also added by the unitarity register wherein the particle register controls the values allowing for multiple different internal bosons. Finally the ancilla qubit $a$ is used to allow for the addition of the two terms in \cref{vertexfactor}. The diagram representation of the gate is seen in \cref{fig:vertexgate} and its full construction is found in Appendix A.

Finally the last gate we list here is the propagator gate $\Prop$. This gate is responsible for implementing the internal bosonic propagator with four-momentum $k_\mu$, which in arbitrary $R_\xi$ gauge reads
\begin{equation}
\Delta_{\mu\nu}(s) = \frac{-i\eta_{\mu\nu}}{s - M_V^2} + \frac{i(1 - \xi_V)k_\mu k_\nu}{(s - M_V^2)(s - \xi_VM_V^2)}
    \label{propagator}
\end{equation}
where $M_V$ is the mass of the boson. To implement this factor into the quantum state we split each term by its scalar and tensorial parts and define the pole factors 
\begin{equation}
\begin{aligned}
    m_0(s) & \equiv \frac{-i}{s - M_V^2}
\andspace \\ m_1(s) & \equiv \frac{i(1 - \xi_V)}{(s - M_V^2)(s - \xi_V M_V^2)}
\end{aligned}
    \label{poles}
\end{equation}
so that $\Delta_{\mu\nu}(s) = m_0(s)\eta_{\mu\nu} + m_1(s)k_\mu k_\nu$. The propagator gate then acts upon the registers $\{i_1, i_2, \U, p, a\}$ by adding these poles through the unitarity register with the value of $M_V$ controlled by the particle register and placing gate representations of the tensors onto the index registers $i_1$ and $i_2$. Again the ancilla qubit is there to allow for the addition of the terms in \cref{propagator}. The circuit representation of the propagator gate is seen in \cref{fig:propgate} whereas the internal details are found in Appendix A.

\begin{figure}[h]
    \centering
    \includegraphics[width=0.9\linewidth]{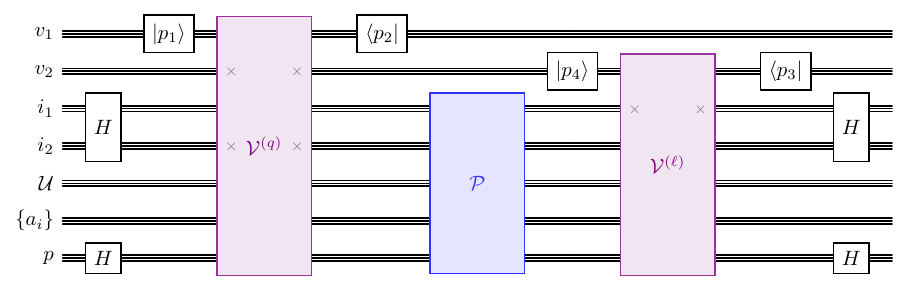}
    \caption{The full quantum circuit for simulating a DY scattering process of the form $q\bar{q} \to \gamma/Z \to \ell^-\ell^+$.}
    \label{fig:DYcircuit}
\end{figure}

Following the standard Feynman rules one can put these primary gates together into one DY circuit as shown in \cref{fig:DYcircuit}. In this circuit it is important to take note that the particle register $p$ is able to encode the internal bosons by mapping the basis states $\{|0\rangle, |1\rangle\}$ to $\{\gamma, Z\}$ respectively, enabling the simultaneous computation of $\M_\gamma$ and $\M_Z$ through the Hadamard transform. Closing the particle register with the last Hadamard then yields $\M_\gamma + \M_Z$ and thus $\big|\M\big|^2 = \big|\M_\gamma + \M_Z\big|^2$ in the readout. For a full derivation of the quantum state of this circuit see Appendix B.

\end{multicols}

\begin{multicols}{2}
    
\vspace*{1mm}
\begin{center}
\subsection{Isolation of interference}    
\end{center}
In any arbitrary scattering process where the matrix element is given by a sum of individual Feynman diagrams $\M = \sum_i\M_i$, the interferences refer to the cross-terms in the squared amplitude:
\begin{equation}
    |\M|^2 = \underbrace{\sum_i|\M_i|^2}_{\text{resonances}} + \underbrace{\sum_{i<j}2\text{Re}\big[\M_i\M_j^*\big]}_{\text{interferences}}.
\end{equation}
Throughout this paper we denote this as $\text{Int}\big[\M_i, \M_j\big]$. The interference terms may contribute to the full cross-section constructively or destructively bringing on completely different outcomes in different energy regimes. Thus studying the interference pattern in a given scattering process may be of valuable interest. This is for instance true in cases of BSM searches where one may want to assess the effects of SM/BSM interference patterns at different energy scales. However, studying the isolated interferences in more complex processes can become rather challenging, much so when the sum of diagrams is large. In this project we will  discuss how to simply make a small basis rotation in the circuit to infer the isolated interference contribution, while not disturbing the full amplitude computation. We stay with the DY process with $\gamma$ and $Z$ as the bosonic propagators where the precise interference is $2\text{Re}\big[\M_\gamma \M_Z^*\big]$ (i.e., within the SM). Looking at the circuit output state $|\psi\rangle$, before the final Hadamard in the particle register, we (up to normalization) have the structure
\begin{equation}
    |\psi\rangle \sim \sum_{V = \gamma, Z}\M _V |V\rangle_p \vac_{\text{work}} \ortho{\vac_{\text{work}}}
\end{equation}
where, again, we have defined $\text{work} \equiv \R\setminus p$. In the above state, everything to the right are terms orthogonal to the vacuum state $\vac_\text{work}$. By projecting to the vacuum state in work using the operator $\mathcal{P}_\Omega = \vac\langle \Omega|_{\text{work}}$ we get the following sub-state $|\psi_p\rangle \equiv\mathcal{P}_\Omega |\psi\rangle$ which has a density matrix given by 
\begin{equation}
\rho_p = |\psi_p\rangle \langle \psi_p| \sim  \sum_{V_1, V_2} \M_{V_1}\M_{V_2}^* |V_1\rangle\langle V_2|.
\end{equation}
It is then easy to see that the information of the interferences is hidden inside the off-diagonal parts of the density matrix $\rho_p$. We can extract this information by rotating and measuring in the $X$-basis where the result will be found in the expectation value $\langle X\rangle$. The rotation is done by the identity $X = HZH$, i.e., we apply a Hadamard gate and then measure in the computational basis. The expectation value is then given by
\begin{equation}
    \langle X \rangle = \langle \psi_p|X|\psi_p\rangle  = \langle \psi_p|HZH|\psi_p\rangle
\end{equation}
with $\langle Z\rangle = (N_0 - N_1)/\text{\#shots}$ where $N_0$ and $N_1$ are the number of $|0\rangle$ and $|1\rangle$ counts respectively during a number of shots of the circuit. This gives exactly the interference by the following:
\begin{equation}
\begin{aligned}
    \langle X\rangle & = \tr\big[\rho_pX\big] \\
    & =  \big(\M_\gamma \M^*_Z + \M_Z\M^*_\gamma\big)  =  \text{Int}\big[M_\gamma,\M_Z^*\big].
\end{aligned}
\end{equation}
Hence we have rotated into the Hadamard basis in the particle register to directly assess the interference. On the upside this does not change our circuit as this rotation is already implemented in our circuit to find the full amplitude squared. Hence we find the full amplitude squared and the isolated interference simultaneously from sampling the circuit with sample size $\#\text{shots}$ and then estimate

\begin{equation}
    \begin{aligned}
         \big|\M_\gamma + \M_Z\big|^2 & \sim \frac{N_0}{\text{\#shots}} \andspace \\
        2\text{Re}\big[\M_\gamma \M_Z^*\big]  & \sim \frac{N_0 - N_1}{\text{\#shots}}.
    \end{aligned}
\end{equation}
This is the primary result of the designed circuit. 

\hspace{3mm}
\begin{centering}
\subsection{Phase space integration}  
\end{centering}
For a fair assessment of the circuit's potential, we would like to aim for a computation of the full integrated cross-section $\sigma$. The path we choose is to discretize the integration over several Phase Space (PS) points, similar to a blanket Monte-Carlo (MC) integration. In this following section we will discuss how to implement a discretized PS into our circuit. As of this point, any arbitrary PS point evaluation of the amplitude squared can be done by defining the spinor and propagator gates in accordance with the values of $\sqrt{s}$ and $\theta$ (the partonic energy and scattering angle respectively). However, in order to reach our goal we must scan the PS in a region of interest. Thus, we are in need of many discrete point evaluations leading to several iterations of the circuit. Instead we utilize superposition to allow for several PS point evaluations with one single circuit. Consider an energy range of interest  
\begin{equation} 
\sqrt{s} \in \Big[\sqrt{s}_\text{min}, \sqrt{s}_\text{max}\Big] \ [\text{GeV}]
\end{equation}
and its discretization into $2^n$ points
\begin{equation}
\begin{aligned}
    & \Big[\sqrt{s}_\text{min}, \sqrt{s}_\text{max}\Big] \\
    & \mapsto \Big\{\sqrt{s}_k \equiv \sqrt{s}_\text{min} + k\Delta s\  \big|\ k = 0, 1, ..., 2^n  -1\Big\}
\end{aligned}
\end{equation}
where $\Delta s = \big|\sqrt{s}_\text{min} -  \sqrt{s}_\text{max}\big|/(2^n-1)$ is the step size. The integer $n$ refers to the number of qubits of an energy quantum register $\mathcal{E}$ which will be used to open a superposition of these energy levels. The point of this method is to map each discrete value $\sqrt{s}_k$ to a basis state in $\mathcal{E}$ with the standard binary integer representation:
\begin{equation}
    \big|\sqrt{s}_k\big\rangle_\E  \in \big\{|0\rangle_\E, |1\rangle_\E, ..., |2^n - 1\rangle_\E \big\}.
\end{equation}
The mapping is straightforward and linear: in \cref{fig:s_disc} we see a schematic view of how the states align with the energy range. 

\begin{figure}[h]
    \centering
    \includegraphics[width = 0.9\linewidth]{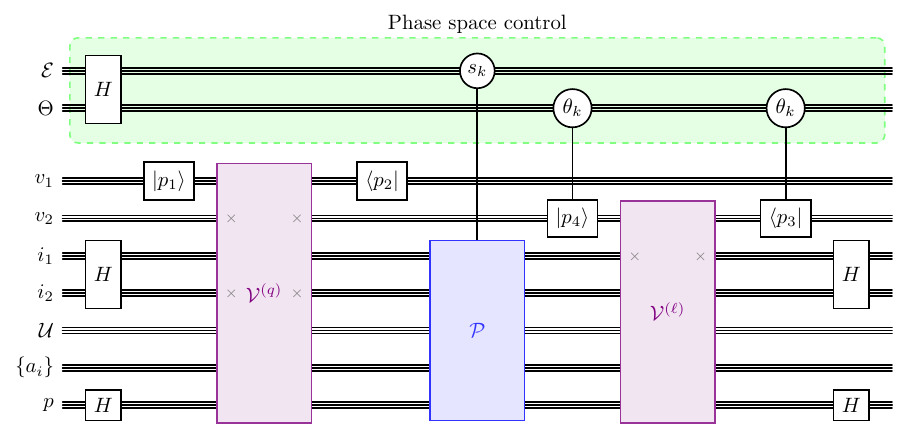}
    \caption{The PS controlled version of the DY circuit with new registers $\E$ and $\Theta$.}
    \label{fig:PS-Drell-Yan_Circuit}
\end{figure}

\begin{Figure}
    \centering
    \includegraphics[width = 1\linewidth]{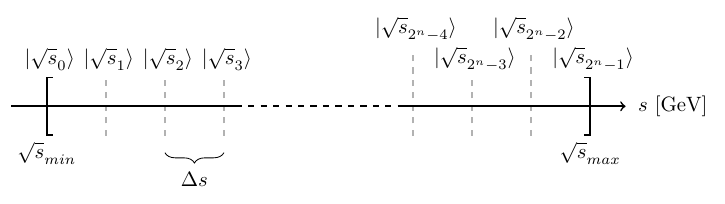}
    \captionof{figure}{Discretization and state encoding of the PS range $\sqrt{s} \in [\sqrt{s}_\text{min}, \sqrt{s}_{\max}]$.}
    \label{fig:s_disc}
    \vspace*{0.45truecm}
\end{Figure}
\noindent The point is that this $\E$ register is able to control the circuit and enforce which value of $\sqrt{s}$ is to be used as input. Then on an abstract level the quantum state in this register is 
\begin{equation}
    |\psi\rangle_\E \sim \sum_{k}\M(s_k)\big|\sqrt{s}_k\big\rangle_\E,
\end{equation}
which provides us with a set of evaluated amplitude squares by projecting to the states of $\E$ to get $\big|\M(s_k)\big|^2 = \big|\langle \sqrt{s}_k|\psi\rangle_\E\big|^2$. In order to achieve this, we note that it is not feasible to use a controlled version of the entire circuit. Rather we only control the gates that are energy dependent: the propagator $\Prop$ and the spinor gates. However, we would prefer to control as few gates as possible so we instead opt for energy-independent spinor inputs. In the spinor gate inputs we rescale each spinor as $u_s(\boldsymbol{p}) \mapsto u_s(\boldsymbol{p})/\sqrt{2E}$ or instead by setting $E = 1/2$ universally so that only the vector structure remains of the spinors. The effect of this on the amplitude level is the rescaling $\M(s) \mapsto \M(s)/(4E^2) = \M(s)/s$, which yields a $1/s^2$ rescaling in the amplitude squared. To account for this one just multiplies with $s^2_k$ in accordance with the states value $\sqrt{s}_k$ in the circuit output.
With the spinor gates completely energy independent one can now control the propagator gate with the energy register to insert different values for the pole terms $m_0(s_k)$ and $m_1(s_k)$ to yield a range of $\M(s_k)$ values. \\ \indent
In a similar manner one desires a discretization of the scattering angle, which is the cross-section integration variable. We only work in the Center-of-momentum (COM) frame and so the $\varphi$ angle is decoupled throughout. The continuous angle range is $\theta \in [0, \pi]$ and so we do a discretization by 
\begin{equation}
    \theta_k \in \big\{ \theta_k \equiv k\Delta\theta \ | \ k = 0, 1, ..., 2^n-1\big\}
\end{equation}
with the step size $\Delta\theta = \pi/(2^n-1)$. Again we implement a quantum register $\Theta$ for which we map its basis states to these discrete points. With this we can control the gates that are scattering angle dependent, i.e., only the outgoing spinor gates. A complete PS controlled version of the DY circuit is shown in \cref{fig:PS-Drell-Yan_Circuit} with the two new registers highlighted. The full state derivation of this circuit is found in Appendix B.

\end{multicols}

\begin{multicols}{2}

Before ending this section we will discuss the normalization factors brought on in this circuit. This gives us the compensation factor $C(s)$ by which one needs to rescale the outputs of the circuit accordingly. There are three factors contributing to the overall normalization of the quantum state. Firstly the Hadamard factors brought on by the opening of superpositions and the ancilla addition operations. From the Hadamard transformations the energy, scattering angle, index and particle registers contribute with factors $\sqrt{2^{n_\E}}, \sqrt{2^{n_\Theta}}, 4$ and $\sqrt{2^{n_p}}$ respectively while the ancilla addition operations yield a total factor of $2^3$. All together this gives a factor $32\times \sqrt{2^{n_\E + n_\Theta + n_p}}$. Secondly we need to take into account the pole normalization factor $\bar{m}(s_k)$ brought on by the propagator gate (see Appendix A for detailed explanation). Lastly we need the spinor gate normalizations, which for $E = 1/2$ yields $|u_s(\boldsymbol{p})| = 1$ $\forall\theta$. Hence we only need to add the aforementioned compensation of $s_k$ and thus the full compensation factor that is used to obtain a correctly scaled output is given by 
\begin{equation}
    C(s_k) = 32 \times   \sqrt{2^{n_\E + n_\Theta + n_p}}\times  s_k\bar{m}(s_k). \label{compensation}
\end{equation}
When receiving the outputs from the circuit, the values for $|\M(s_i, \theta_j)|^2$ are found by scaling the outputs by $C(s_i)^2$. \\ \indent
In conclusion, we have developed a complete quantum circuit for simulating two-body DY scattering for an arbitrary number of internal bosons in a discretized PS built from $n_s$ energy points and $n_\theta$ angle points. The full circuit is built from nine registers where the number of qubits needed for each register is presented in \cref{tab:qubit_table}. A conclusion that can be drawn from the table is the fact that the qubit scaling comes from the PS registers, assuming a small number of internal bosons $n_p$, which shows a logarithmic scaling with the resolution of the PS discretization.

\end{multicols}

\begin{table}[h]
    \centering
    \begin{tabular}{||c||c|c|c|c|c|c|c|c|c||}
    \hline
    Register &  $\E$  & $\Theta$ & $v_1$ & $v_2$ & $i_1$ & $i_2$ & $\U$ & $\{a_i\}$  & $p$ \\ \hline
    Number of qubits & $\lceil \log_2(n_s) \rceil$ & $\lceil\log_2(n_\theta)\rceil$ & $2$ & $2$ & $2$ & $2$ & $3$ & $3$ & $\lceil\log_2(n_p)\rceil$ \\\hline
    \end{tabular}
    \caption{Number of qubits needed for each register in the circuit of \cref{fig:PS-Drell-Yan_Circuit}.}
    \label{tab:qubit_table}
\end{table}


\begin{multicols}{2}

\begin{figure}[H]
    \centering
    \subfigure[Energy distribution.]{\includegraphics[width = 0.46\textwidth]{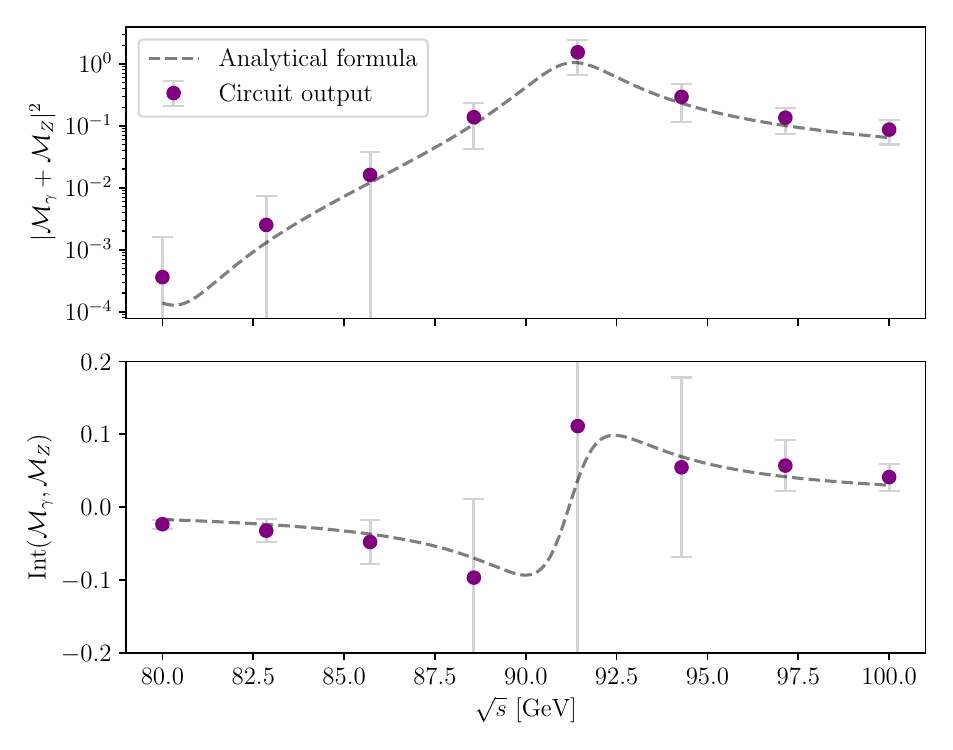}\label{fig:splice_s}}
    \subfigure[Angular distribution.]{\includegraphics[width = 0.44\textwidth]{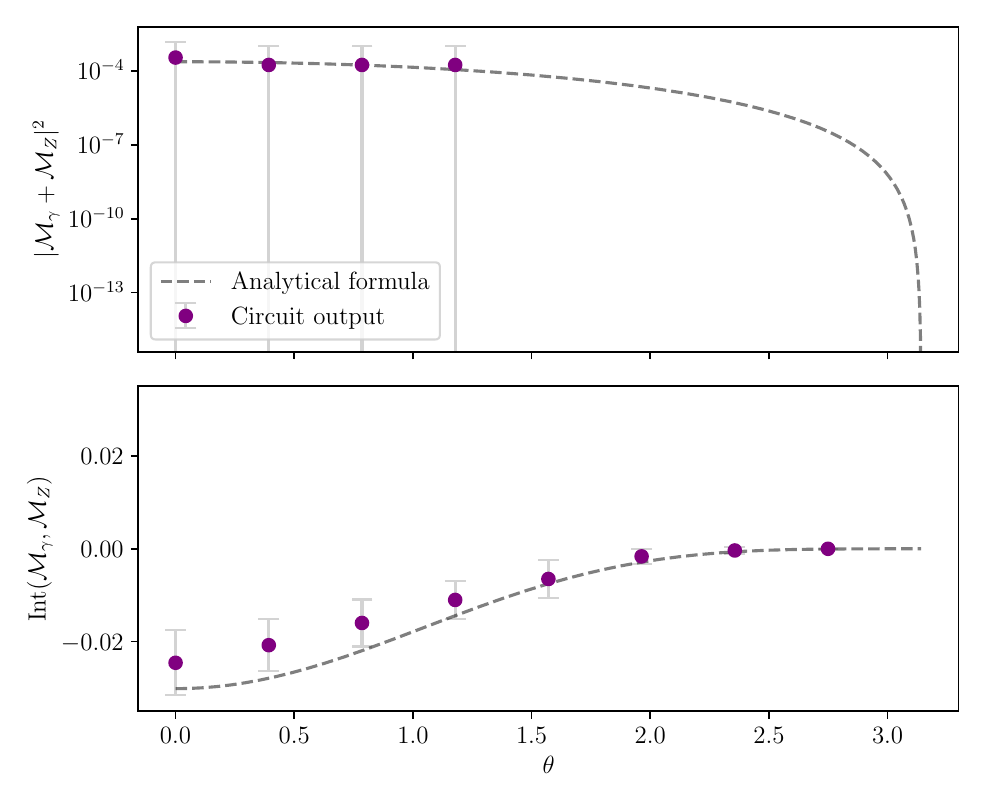}\label{fig:splice_theta}}
    \caption{Circuit output for DY scattering with (a) fixed $\theta = \pi/3$ and (b) fixed $\sqrt{s} = 80$ GeV at the amplitude level with full amplitude squared (top) and the associated interference pattern (below). In both simulations the circuit was run for $2\times 10^7$ shots and 25 batches where the presented errors are the standard deviations of the batches and the points are the batch averages. The true value was computed from the analytical expressions for $\M_\gamma$ and $\M_Z$ in \crefrange{A:Mgamma}{A:MZ}. For viewing convenience the interference plot has been zoomed in, the large uncertainty at the resonance peak reach up to $\sim 0.8$.}
\end{figure}

\begin{center}
    \section{Results}
\end{center}
In this section we show different results from the circuit in \cref{fig:PS-Drell-Yan_Circuit} with associated expectations from true analytical expressions of \crefrange{A:Mgamma}{A:MZ} for comparisons. Throughout this section we perform a test on the DY process $u\bar{u} \to \gamma/Z \to \mu^-\mu^+$ where the helicity configuration is $(+,-,+,-)$. We also work in Feynman gauge $\xi_V = 1$ (as mentioned) where the double pole in the propagator vanishes, simplifying the implemented circuit. The circuit we have designed is presumably outside the scope of Noisy Intermediate-Scale Quantum (NISQ) era quantum computing and so we do not attempt to test it on real quantum devices as it would only infer unreliable outputs. Instead we opt to simulate the circuit using Qiskit \cite{javadiabhari:2024} through classical means. In order to efficiently store the data of the circuit simulation we note that all the output strings can either be kept if the work registers are in the vacuum (hit) or discarded otherwise (miss). Since a large number of such outputs will be discarded there is no need to keep all the individual output strings that are misses and so we use an ancilla register of one qubit that records if the output state is a hit or miss. This makes it so that a much smaller number of qubits need to be measured and fewer output states have to be saved in the classical memory, saving both time and resources. More details of this can be found in Appendix A.
\vspace{3mm}
\begin{centering}
    \subsection{Amplitude level} 
\end{centering}
\vspace{-3.5mm}The first test involves a stationary scattering angle at $\theta = \pi/3$ with the energy range $\sqrt{s}\in[80, 100]$ discretized by $8$ points. This requires $3$ qubits for the $\E$ register in order to encode the $8$ energy points. The circuit was run with $2\times 10^7$ shots for $25$ batches and the results are shown in \cref{fig:splice_s} against the expected outcome. The circuit outputs are the averages of the batches and the error bars are the standard deviations of the batches. 

Similarly we also run a test for the angular distribution of the amplitude. We set $\sqrt{s} = 80$ [GeV] and discretize the scattering angle range by $8$ points, again requiring $3$ qubits for the $\Theta$ register. The results are shown in \cref{fig:splice_theta}. The four missing points of the top figure in \cref{fig:splice_theta} were not found by the circuit as their probability amplitudes were rapidly vanishing for increasing $\theta$.

\end{multicols}

\begin{figure}[h]
    \centering
    \includegraphics[width=0.9\linewidth]{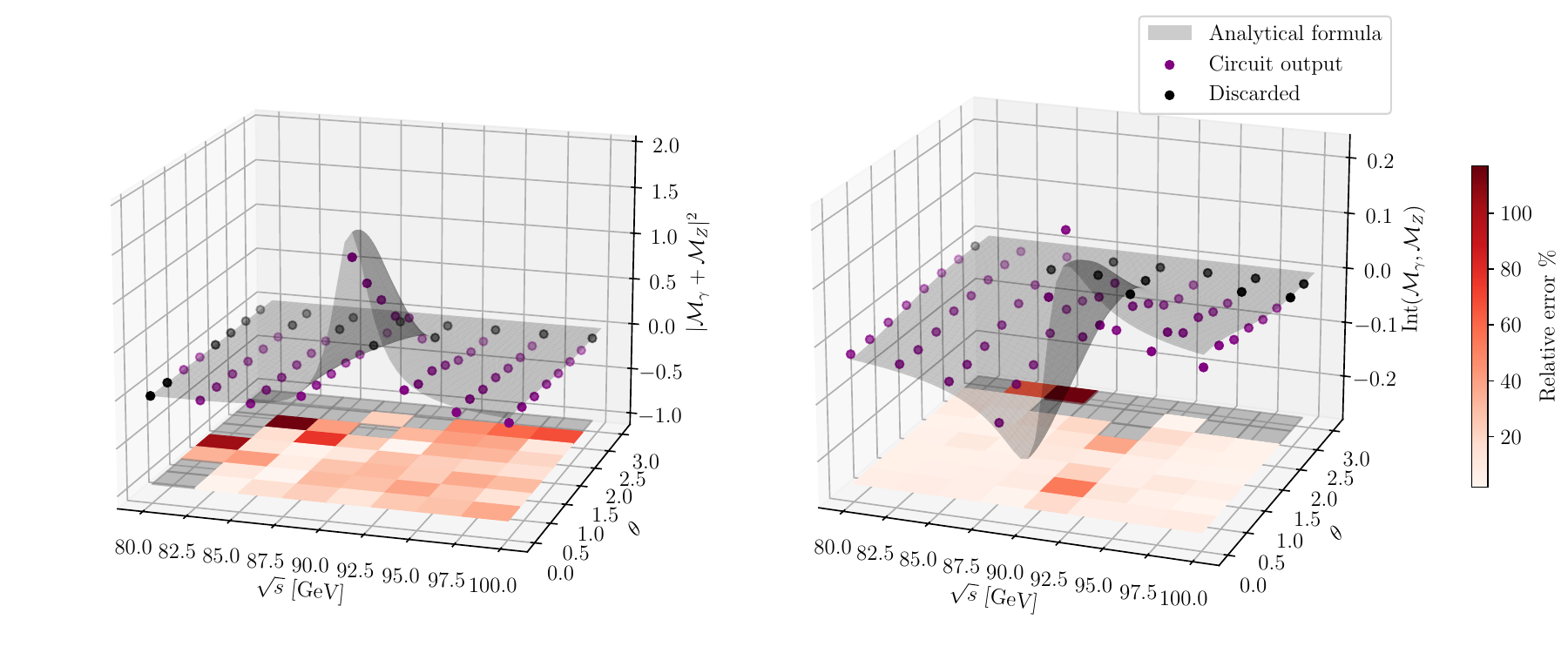}
    \caption{Discrete amplitude level circuit outputs in the grid $[80, 100]\times [0, \pi]$ with full amplitudes (left) and interference pattern (right). The relative error matrices below the surfaces indicate how far off each point is to the analytical formula and the matrices can also be viewed in \cref{fig:Error_matrices}.}
    \label{fig:Surfaces}
\end{figure}

\begin{figure}[h]
    \centering
    \includegraphics[width=0.75\linewidth]{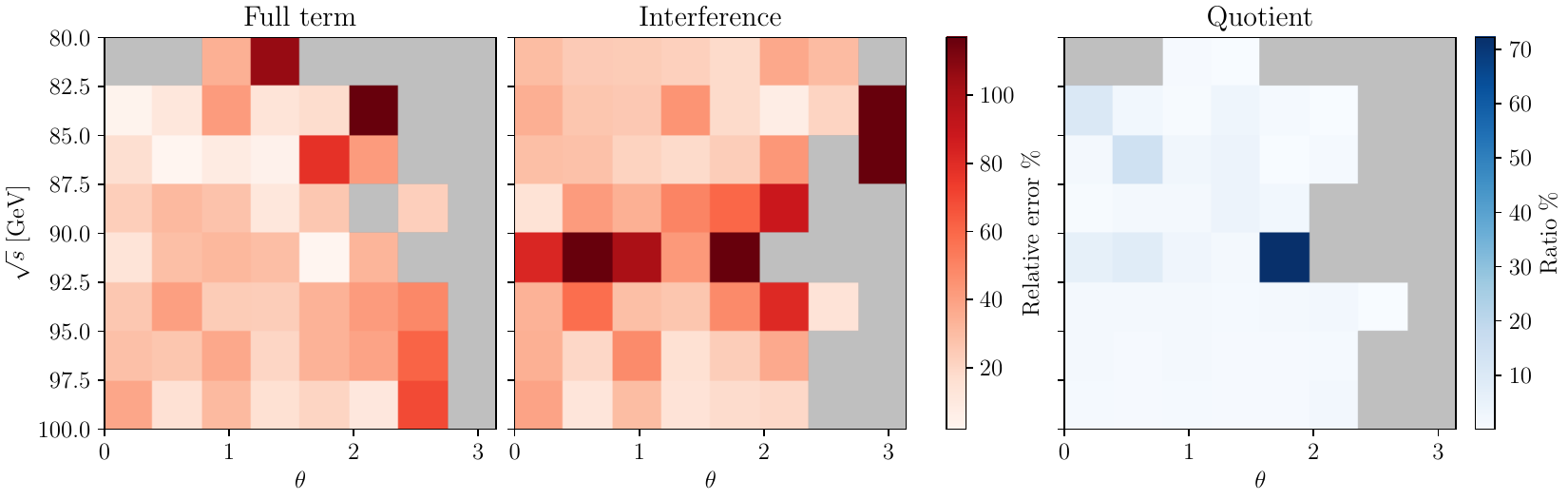}
    \caption{The relative error matrices of \cref{fig:Surfaces} and their quotient. The average errors of the two matrices are $31.117\%$ and $55.305\%$ respectively while the quotient matrix has the average value $3.826\%$.}
    \label{fig:Error_matrices}
\end{figure}

\begin{multicols}{2}
To reach the cross-section we want to compute the amplitude for a grid of $\sqrt{s}$ and $\theta$ values. Our final test at the amplitude level is thus a discretization of the grid $[80, 100] \times [0, \pi]$ by $8\times 8$ points using $3$ qubits each for the $\E$ and $\Theta$ registers. The results from $3\times 10^7$ shots and $50$ batches are shown in \cref{fig:Surfaces}. We again show the averages of the batches but without the standard deviation. Instead we show associated relative errors from the true analytical expression, i.e., the shaded surfaces. These relative error matrices are also presented in \cref{fig:Error_matrices} together with their quotient to get an assessment of how accurate the interference outputs are in comparison with the full term outputs. The discarded points of \crefrange{fig:Surfaces}{fig:Error_matrices} are outputs that are deemed to be outliers or are points that were not found in the outputs of the circuit. The majority of these discarded points are ones with $\big|\M\big|^2 \approx 0$ and so are internally very difficult to sample. 

\end{multicols}

\begin{multicols}{2}
\begin{center}
    \subsection{Cross-section integration} 
\end{center}
Given the outputs of the previous section we now integrate over the scattering angles to reach the final cross-section $\sigma(u\bar{u} \to \gamma/ Z \to \mu^-\mu^+)$. Recall that we now have a set of amplitude outputs for a discrete phase space $\big\{\sqrt{s}_i, \theta_j\big\}$ given by the circuit as seen in \cref{fig:Surfaces}. To reach an approximate cross-section integration we discretize the full integral by summing over all the $\theta_j$ outputs for each $\sqrt{s}_i$. The discretization is given by 
\begin{equation}
\begin{aligned}
    \sigma(s_i) & = \frac{1}{3}\frac{1}{64\pi^2 s_i}\int \text{d}\cos\theta\int \text{d}\varphi \big|\M(s_i, \theta)\big|^2  \\
    & \approx \frac{1}{3}\frac{1}{64\pi^2 s_i}2\pi\sum_{\theta_k}\Delta\theta \sin(\theta_k) \big|\M(s_i, \theta_k)\big|^2,
\end{aligned} 
\label{cross-section discretization}
\end{equation}
where in the COM frame the $\varphi$ angle only contributes with a $2\pi$ factor (as intimated). Recall that we are looking at one helicity configuration, but summing and averaging over the different quark colors, hence the factor $1/3$ up front. The same method is performed for the interference cross-section $\sigma_\text{int}$ where we replace $\big|\M(s_i, \theta_k)\big|^2$ with $\text{Int}\big[\M_\gamma(s_i, \theta_k), \M_Z(s_i, \theta_k)\big]$ in \cref{cross-section discretization}. To estimate the errors of the discretization we use standard error propagation. The standard deviations $\sigma_{ik}$ of the batches are used as errors and so the uncertainty of each $\sqrt{s}_i$ point is given by 
\begin{equation}
    \delta\sigma(s_i) = \frac{1}{3}\frac{2\pi}{64\pi^2 s_i}\sqrt{\sum_{k}\big(\Delta\theta \sin(\theta_k)\big)^2 \sigma_{ik}^2}.
    \label{cross-section err}
\end{equation}
We aim to compare this discrete integration with standard MC  integration and thus perform a computation of these same quantities in \texttt{MadGraph} \cite{Alwall:2014}. Here \texttt{MadGraph} performed flat PS generation, i.e., no importance sampling, to match the way that the quantum circuit generates an equal superposition of the PS points. Of course one could implement importance sampling into the quantum circuit by preparing a weighted superposition in the PS registers, although this is not taken into account into this work.

In \cref{fig:cross_section} we show our final results of the integrated cross-sections  for both the full amplitude $\sigma$ and interference $\sigma_\text{int}$ using the output points of \cref{fig:Surfaces}. Here the circuit and \texttt{MadGraph} outputs, including their respective uncertainties, are compared to a numerical integration done over the analytical expressions of the amplitudes in \crefrange{A:Mgamma}{A:MZ} as benchmark values. In the lowest plot we show the uncertainty quotient $\delta\sigma/\delta\sigma_\text{int}$ to get an idea of how accurate either method is at estimating both quantities simultaneously, or from the same respective sample size. 

\end{multicols}

\begin{Figure}
    \centering
    \includegraphics[width = 0.7\linewidth]{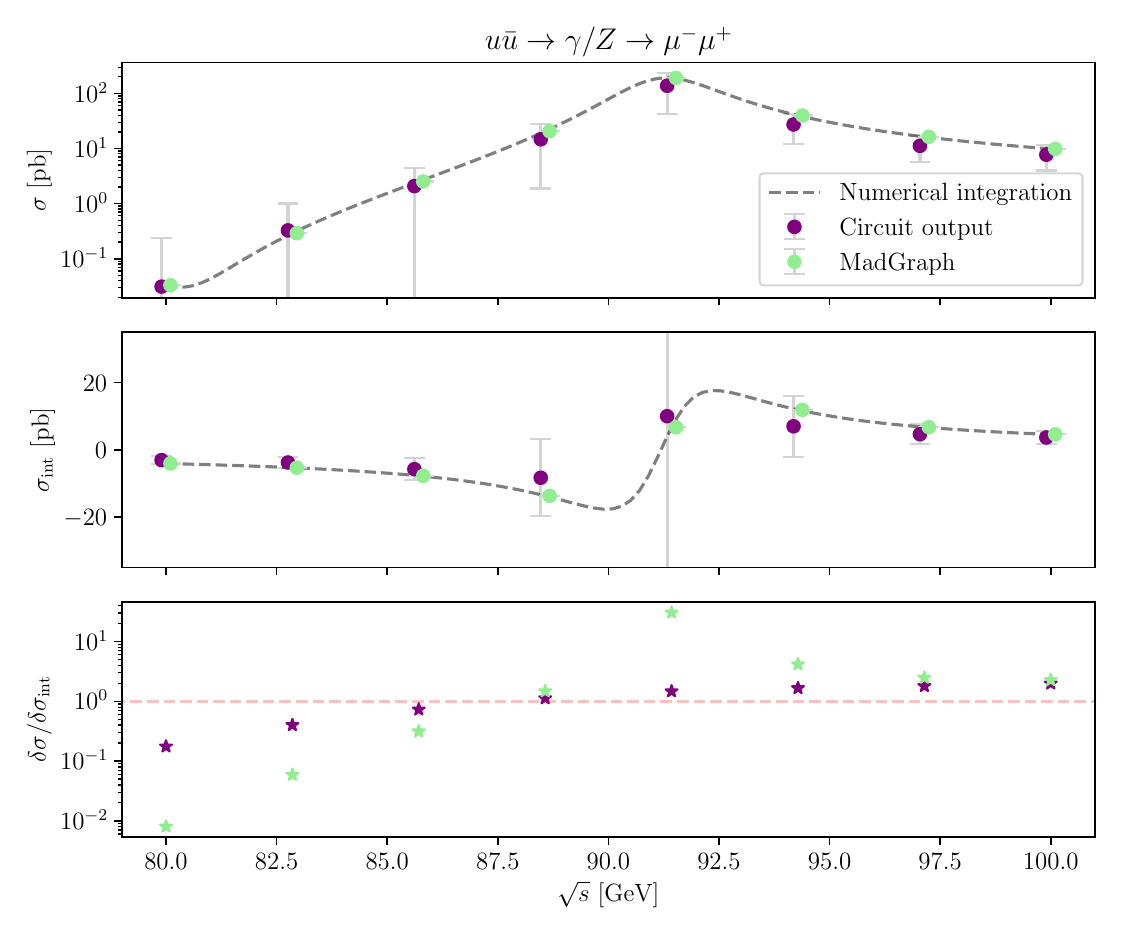}
    \captionof{figure}{Integrated cross-section results from the circuit and \texttt{MadGraph} compared to numerical integration including both the full amplitude $\sigma$ and the interference $\sigma_\text{int}$. The lowest plot shows the uncertainty quotient $\delta\sigma/\delta\sigma_\text{int}$ for both methods where the red dashed line indicates where $\delta\sigma = \delta\sigma_\text{int}$.}
    \label{fig:cross_section}
\end{Figure}

\begin{multicols}{2}
\begin{center}
    \section{Discussion and summary}
\end{center}
The quantum circuit developed in this work provides a complete gate-based formulation of
partonic DY scattering. By translating spinors, vertices, propagators and index
contractions into unitary operations, the circuit embeds the full amplitude into a quantum
state whose measurement yields both the total contribution and, crucially
for our purposes, the isolated interference
between $\gamma$ and $Z$ exchange. Indeed, such a method aims to "\textit{sum and square
their contributions together while also simultaneously isolating their  interference}", a
feature that is essential for many realistic collider studies.

Simulations show that the circuit reproduces analytical expectations with good accuracy
across the tested phase space. The full amplitude remains stable, while the isolated
interference term becomes increasingly difficult to resolve near the $Z$ resonance, where the
interference naturally approaches zero and the corresponding quantum amplitudes become
extremely small. Despite this challenge, the circuit successfully extracts both quantities within
the expected precision, the computation of the total amplitude being comparable to the \texttt{MadGraph} output and that of the interference outperforming it (for a comparable number of tested phase space points). Taking a closer look at the lowest plot of \cref{fig:cross_section} we note that this data answers the question "\textit{how accurately can a method estimate both types of cross-sections simultaneously}?". In the case of \texttt{MadGraph}, the word \textit{simultaneous} implies that the computations were done from the same MC samples. The red line in the figure highlights the spot where the uncertainties are equal $\delta\sigma = \delta\sigma_\text{int}$, i.e., an ideal situation where the method is equally certain about both quantities. Our main conclusion from this graph is thus that even if \texttt{MadGraph} is considerably more certain at each individual point, the uncertainty quotient for the circuit is closer to the ideal line throughout the studied energy range. It is thus shown to reliably estimate both cross-sections simultaneously. This implies that quantum methods of this kind could be used more efficiently when estimating multiple correlated quantities.

The main practical limitation of the QC approach developed here is the low probability of measuring the vacuum state that
encodes the physical amplitude, which necessitates a large number of samples of the circuit. Addressing this
will likely require amplitude amplification or alternative readout strategies. 

Finally, future extensions to this work are quite clear. Firstly one needs to attack the aforementioned sample bottleneck problem, but beyond that some prominent ideas are: 
use of importance sampling to resolve the missing points of \cref{fig:splice_theta}, inclusion of helicity sums and PDFs as well as extensions to more complex scattering processes where interference patterns are classically difficult to compute. Additionally, one can easily take this given circuit and extend it to BSM analyses for $Z^\prime$ bosons by including an extra qubit in the particle register. It is then straightforward to compute the SM/BSM interference by performing a similar basis rotation as in this work. 

\vspace{3mm}
\begin{centering}
    \subsection*{Acknowledgments}
\end{centering}
The authors  thank IBM for the open-source quantum computing platform IBM Quantum and their work on the Python module Qiskit, making this project possible. 
S.~M. is supported in part through the NExT Institute and STFC Consolidated Grant ST/X000583 /1.
T.V.~is supported by the Swedish Research Council under contract number VR:2023-00221. The computations were enabled by resources within the project UPPMAX 2025/2-312 provided by the National Academic Infrastructure for Supercomputing in Sweden (NAISS). 


\end{multicols}

\section*{Appendix A: Circuits and formulas}    

\setcounter{equation}{0}
\renewcommand{\theequation}{A.\arabic{equation}}
\setcounter{figure}{0}
\renewcommand{\thefigure}{A.\arabic{figure}}

\textbf{Operator addition} \\
In many contexts throughout this paper we are determined to implement a combination of unitary operators $A + B$ in which the addition (or subtraction) does not necessarily need to be unitary. To attack this dilemma we introduce an ancilla qubit. Consider the initial state $\vac_t|0\rangle_a$ with a target register $t$ and ancilla qubit $a$. Add a Hadamard to the ancilla qubit and then use controlled operators $C_{|0\rangle}\big[A\big]$ and $C_{|1\rangle}\big[B\big]$ and one gets: 
\begin{equation}
    \vac_t|0\rangle_a \to \frac{1}{\sqrt{2}}\bigg(A\vac_t|0\rangle_a + B\vac_t|1\rangle_a\bigg).
\end{equation}
Applying the Hadamard once again gives the summation of $A$ and $B$ in front of $|0\rangle$ and the subtraction in front of $|1\rangle$:
\begin{equation}
\begin{aligned}
     \frac{1}{\sqrt{2}}\bigg(A\vac_t|0\rangle_a + B\vac_t|1\rangle_a\bigg)   \to \frac{1}{2}\bigg[\big(A + B\big)\vac_t|0\rangle_a \textcolor{gray}{+ \big(A - B\big)\vac_t|1\rangle_a}\bigg].\\
\end{aligned}
\end{equation}
Diagrammatically one can see this in \cref{fig:Op_sum}.

\begin{Figure}
    \centering
    \includegraphics[width = 0.4\textwidth]{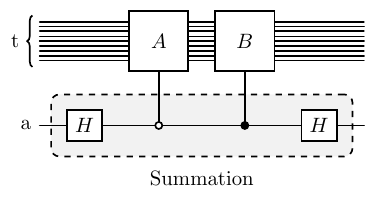}
    \captionof{figure}{Summation of the operators $A$ and $B$ through ancilla control.}
    \label{fig:Op_sum}
\end{Figure}

\noindent\textbf{Quantum gates}\\
In this  section of the appendix we will discuss the concrete implementations of the different quantum gates used for the development of the DY scattering circuit. These include for instance quantum gate representations of the Dirac matrices $\gamma^i$ with $i = 0, ..., 5$, a quantum gate version of the Minkowski metric $\eta = \text{diag}(-1, +1, +1, +1)$ for contracting the indices of the diagrams and also more convoluted gates for the vertices and propagators of the diagrams. \\

\begin{figure}[h]
\centering
    \subfigure[]{\includegraphics[]{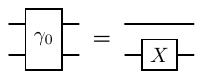}}
    \subfigure[]{\includegraphics[]{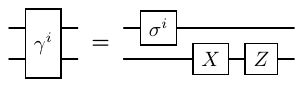}}
    \subfigure[]{\includegraphics[]{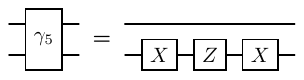}}
\caption{Circuit diagrams for the Dirac matrices represented in  \cref{Dirac_matrices}.}
\label{fig:Dirac circuits}
\end{figure}

\noindent\textbf{Dirac gates}\\
In the following we use the chiral Weyl basis where the unitary Dirac matrices read
\begin{align}
    \gamma^0 & = \sigma_1\otimes\mathds{1}, \hspace{3mm}
    \gamma^i = i\sigma_2\otimes \sigma_i = (\sigma_3\sigma_1)\otimes \sigma_i \andspace
    \gamma^5 = -\sigma_3\otimes\mathds{1}  = (\sigma_1\sigma_3\sigma_1)\otimes \mathds{1}
    \label{Dirac_matrices}
\end{align}
in addition to the relation $\beta = \gamma^0$. From this structure the Dirac matrices are straight-forward to implement as two-qubit gates form the fact that the Pauli matrices are simply the ordinary $X$, $Y$ and $Z$ standard gates. The circuit diagram for these can be seen in \cref{fig:Dirac circuits}.

\begin{figure}[h]
    \centering
    \includegraphics[width = 0.6\linewidth]{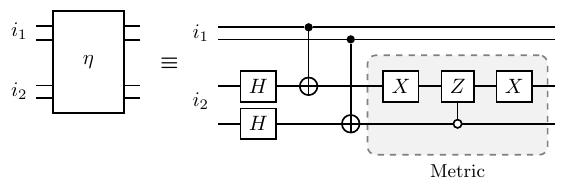}
    \caption{Circuit representation of the $\eta$ gate.}
    \label{fig:eta_circ}
\end{figure}

\noindent\textbf{Minkowski metric gate}\\
To implement the mostly-plus Minkowski metric $\eta_{\mu\nu} = \text{diag}(-1, +1, +1, +1)$ in terms of a quantum operator we first define an index register $i$. This index register needs to encode four values: one temporal $\mu = 0$ and three spatial $\mu = 1, 2, 3$ and thus is composed of two qubits where a state is represented by the binary representation
\begin{equation}
\begin{aligned}
|\mu\rangle_i & \in \{|0\rangle, |1\rangle, |2\rangle, |3\rangle\}  = \{|00\rangle, |01\rangle, |10\rangle, |11\rangle\}.
\end{aligned}
\end{equation}
We now want to assign the metric to these states, i.e., flip the sign if $\mu = 0$ and leave it be otherwise. One can check that the two-qubit operator $\big(\mathds{1}\otimes X\big) C_{|0\rangle}\big[Z\big]\big(\mathds{1}\otimes X\big)$, where $C_{|0\rangle}\big[Z\big]$ is a controlled-$Z$ gate, satisfies this. Now we can define the action of the Minkowski gate $\eta$ on two index registers in the equal superposition $\sum_{\mu\nu}|\mu\nu\rangle_{i_1i_2}/4$ to be
\begin{equation}
\begin{aligned}
     \eta\bigg[\frac{1}{4}\sum_{\mu,\nu}|\mu\nu\rangle_{i_1i_2}\bigg]  = \frac{1}{2}\sum_{\mu,\nu}\eta_{\mu\nu}|\mu\nu\rangle_{i_1i_2}  = \frac{1}{2}\bigg(-|00\rangle + |11\rangle + |22\rangle + |33\rangle\bigg)_{i_1i_2}. \label{eta_action}
\end{aligned}
\end{equation}
The action is such that the second index register $i_2$ is brought back to the vacuum $\sum_\nu|\nu\rangle_{i_2} \mapsto \vac_{i_2}$ by a Hadamard transform where a copy of the state of $i_1$ is made with CNOT gates $\sum_\mu|\mu\rangle_{i_1}\vac_{i_2} \mapsto \sum_\mu|\mu\rangle_{i_1}|\mu\rangle_{i_2}$ and finally the metric signature is added only to $i_2$ with the aforementioned signature operator. The full circuit diagram for the $\eta$ gate is depicted in \cref{fig:eta_circ}.

\begin{figure}
    \centering
    \subfigure[]{\includegraphics[width=0.5\linewidth]{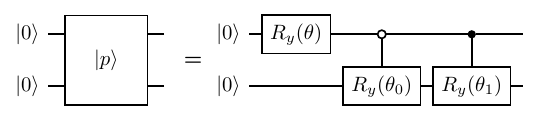}\label{fig:pcircuit}}
    \subfigure[]{\includegraphics[width=0.39\linewidth]{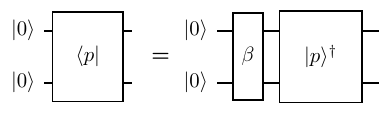}\label{fig:Ubar}}
    \caption{Circuit diagrams the spinor gates (a) $U_p$ which produces the state $|p\rangle$ in \cref{spinorstate} and (b) $\bar{U}_p$ defined in \cref{Up_bar}.}
    \label{fig:spinor_gates_decomp}
\end{figure}

\vspace*{0.5truecm}
\noindent\textbf{Spinor gates}\\
Now we turn to the implementation of the spinor gates. We aim to find quantum gates $U_p$ and $\bar{U}_p$ that implements the components of any arbitrary four-component spinor $u_s(\boldsymbol{p}) = (u_0, u_1, u_2, u_3)^T$ and its conjugate counterpart $\bar{u}_s(\boldsymbol{p}) \equiv u^\dagger_s(\boldsymbol{p})\beta = (u_2^*,  u_3^*, u_0^*,  u_1^*)$ respectively where $\beta = \gamma_0$. Consider this arbitrary spinor as a two-qubit state defined as   
\begin{equation}
    |p\rangle \equiv \frac{1}{|u_s(\boldsymbol{p})|}\sum_{i = 0}^3u_i|i\rangle = \sum_{i = 0}^3\hat{u}_i|i\rangle \label{spinorstate}
\end{equation}
where we have the normalized spinor components $\hat{u}_i$ as probability amplitudes of the orthogonal states $|i\rangle \in \big\{|00\rangle, ..., |11\rangle\big\}$. This qubit state can also be written as $|p\rangle = u_s(\boldsymbol{p})/|u_s(\boldsymbol{p})|$. To be able to initialize this state we expand it and factorize to get 
\begin{equation}
\begin{aligned}
    |p\rangle  & = \hat{u}_0|00\rangle + \hat{u}_1|01\rangle  + \hat{u}_2|10\rangle  + \hat{u}_3|11\rangle  \\
    & = |0\rangle \otimes \big(\hat{u}_0|0\rangle + \hat{u}_1|1\rangle\big)  + |1\rangle \otimes \big(\hat{u}_2|0\rangle + \hat{u}_3|1\rangle\big)\\
    & = \alpha |0\rangle|\varphi_0\rangle + \beta |1\rangle|\varphi_1\rangle
\end{aligned}
\end{equation}
where we have defined the single-qubit states
\begin{equation}
\begin{aligned}
    |\varphi_0\rangle  = \frac{1}{\alpha}\big(\hat{u}_0|0\rangle + \hat{u}_1|1\rangle\big) \andspace |\varphi_1\rangle  = \frac{1}{\beta}\big(\hat{u}_2|0\rangle + \hat{u}_3|1\rangle\big) \label{phis}
\end{aligned}
\end{equation}
and the amplitudes 
\begin{align}
    \alpha  = \sqrt{|\hat{u}_0|^2 + |\hat{u}_1|^2} \andspace
    \beta  = \sqrt{|\hat{u}_2|^2 + |\hat{u}_3|^2}
\end{align}
such that $|\alpha|^2 + |\beta|^2 = 1$. The implementation of $|p\rangle$ on a $2$-qubit circuit becomes simple then by letting the first qubit be a control qubit that conditionally prepares $|\varphi_0\rangle$ if $|0\rangle$ and $|\varphi_1\rangle$ if $|1\rangle$ onto the other qubit. First prepare the single-qubit state $\alpha|0\rangle  + \beta|1\rangle$ with the $R_y(\theta)$ gate that has the action 
\begin{equation}
    R_y(\theta)|0\rangle = \cos(\theta/2)|0\rangle + \sin(\theta/2)|1\rangle
\end{equation}
on the vacuum by rotating by an angle $\theta = 2\cos^{-1}(\alpha)$. Afterwards one uses controlled $R_y(\theta_i)$ rotations onto the second qubit with the rotation angles $\theta_0 = 2\cos^{-1}(u_0/\alpha)$ and $\theta_1 = 2\cos^{-1}(u_2/\beta)$ which prepares the states $|\varphi_0\rangle$ and $|\varphi_1\rangle$ respectively. This procedure can be seen in the circuit diagram representation in \cref{fig:pcircuit} and is the gate that we denote by $U_p$ that has the desired action $U_p|\Omega\rangle = |p\rangle$. We also define the barred gate seen in \cref{fig:Ubar} as
\begin{equation}
    \bar{U}_p \equiv U^\dagger_p\beta
    \label{Up_bar}
\end{equation}
where $\beta$ is a quantum gate that has the same structure as the $\gamma_0$ gate. From this gate one can construct the conjugate state $\langle\bar{p}| = \langle \Omega|U^\dagger_p\beta$.

A small caveat that we need to address in this implementation is that since the states $|\varphi_0\rangle$ and $|\varphi_1\rangle$ are prepared by an angle dependent on $u_0$ and $u_2$ respectively, the information of the signs of $u_1$ and $u_3$ will be lost. To fix this we simply add a controlled-$Z$ gate onto the second qubit that flips the sign of $|1\rangle$ in \cref{phis} if either $u_1$ or $u_3$ are negative. \\

\begin{figure}[h]
    \centering
    \includegraphics[width=0.75\linewidth]{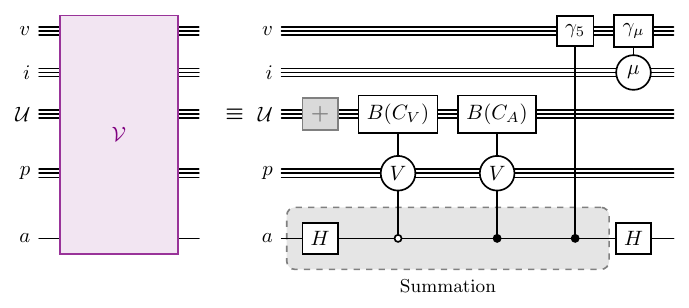}
    \caption{Circuit diagram decomposition for the vertex gate $\mathcal{V}$ that implements the factor in \cref{A:vertex factor}.}
    \label{fig:Vgate_decomp}
\end{figure}

\begin{figure}[h]
    \centering
    \includegraphics[width=0.9\linewidth]{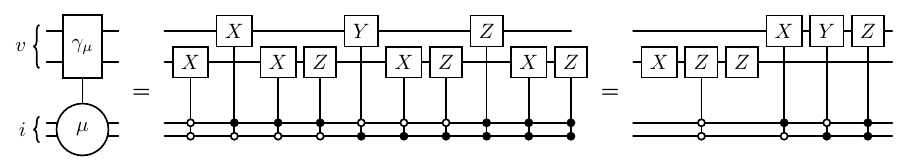}
    \caption{Gate number reduction of multi-controlled Dirac gate.}
    \label{A:multi-gamma}
\end{figure}

\noindent\textbf{Vertex gate}\\
Here we look at the exact structure of the vertex gate. First recall the $Vf\bar{f}$ vertex factor for any EW boson $V = A, Z, W^\pm$ and a fermion pair $f \bar{f}$;
\begin{equation}
    \V_\mu^{(V)} = C_\mathcal{V}^{(V)}\gamma_\mu + C_\mathcal{A}^{(V)}\gamma_\mu\gamma_5
    \label{A:vertex factor}
\end{equation}
with vector and axial-vector couplings $C_\V^{(V)}$ and $C_\A^{(V)}$. These couplings are independent of the kinematics and can be written in terms of quantum numbers via
\begin{equation}
\begin{aligned}
         \boxed{V = \gamma:} & \hspace{3mm}
         C_\V^{(\gamma)} = -ieQ \andspace C_\A^{(\gamma)} = 0 \\
     \boxed{V = Z:} & \hspace{3mm}
      C_\V^{(Z)}  = \frac{ie}{2c_w}\Bigg(\frac{I_3}{s_w} -2s_wQ\Bigg) \andspace 
C_\A^{(Z)}  = -\frac{ie}{2c_ws_w}I_3
\end{aligned}
\label{A:couplings}
\end{equation}
where $c_w \equiv \cos(\theta_w)$, $s_w \equiv \sin(\theta_w)$ are the weak-mixing terms with angle $\theta_w = \arcsin(0.47143025548)$ and $I_3$ and $Q$ are the third isospin component and charge respectively. Our first objective is to use the Dirac gates to construct a vertex gate $\V$ which provides us with this vertex factor to our quantum state. For this we will first need the increment and value setting gates $U_+$ and $B(\alpha)$ developed in \cite{Chawdhry:2023jks} to apply the couplings, hence we will need a unitarity register $\U$ (see appendix in Ref. \cite{Bashore:2025uwb} for a quick recap of this method). We will also need some two-qubit register for which the Dirac gates can be operated upon, which we call a vertex register and denote it by $v$. Lastly we need an ancilla qubit to perform the addition of the two terms in the vertex factor as explained at the beginning of this appendix where its $|0\rangle$ state connects to the vector term and its $|1\rangle$ state connects to the axial-vector term. To enable the possibility of having multiple types of bosons, $V = V_1, V_2, ...V_n$ let us also use a particle register $p$ which controls which couplings are inserted in the value setting gates. Collecting all the needed registers we have $\{v, i, \U, p, a\}$ for which $\V$ acts upon. \\ \indent
The vertex gate works as follows; first we open up the superposition in the ancilla qubit $\vac_a \mapsto (|0\rangle + |1\rangle)_a/\sqrt{2}$ so that it can control the value setting gates. Secondly, together with an increment gate, the controlled value setting gates $\prod_{V_i}C_{|0\rangle}\big[B(C_\V^{(V_i)})\big]C_{|1\rangle}\big[B(C_\A^{(V_i)})\big]$ read the state of the ancilla qubit and the particle register and append the corresponding coupling $C_{\V/\A}^{(V_i)}$. We also want to append $\gamma_5$ only to the axial-vector and so we use the controlled version $C_{|1\rangle}\big[\gamma_5\big]$ and then a set of Dirac gates controlled by the index register $\prod_\mu C_{|\mu\rangle}\big[\gamma_\mu\big]$. To close everything we use another Hadamard in the ancilla such that the vector and axial-vector factors are added together,
\begin{equation}
\begin{aligned}
     \sum_{V_i}\bigg(C_\V^{(V_i)}\gamma_\mu \vac_v|0\rangle_a +C_\A^{(V_i)}\gamma_\mu\gamma_5\vac_v|1\rangle_a \bigg)|V_i\rangle_p &  \xrightarrow[]{H} \sum_{V_i}\bigg(C_\V^{(V_i)}\gamma_\mu +C_\A^{(V_i)}\gamma_\mu\gamma_5 \bigg)\vac_v|0\rangle_a|V_i\rangle_p \\
    & \textcolor{gray}{+ \sum_{V_i}\bigg(C_\V^{(V_i)}\gamma_\mu  - C_\A^{(V_i)}\gamma_\mu\gamma_5 \bigg)\vac_v|1\rangle_a|V_i\rangle_p}
\end{aligned}
\end{equation}
and thus end up with the corresponding vertex factor in front of the $|0\rangle_a$ state where the faded term associated with $|1\rangle_a$ is irrelevant for our case. A circuit diagram of the vertex gate is seen in \cref{fig:Vgate_decomp}.
The multi-controlled Dirac gate $\prod_\mu C_{|\mu\rangle}\big[\gamma_\mu\big]$ present in \cref{fig:Vgate_decomp} can be simplified to reduce the total number of controlled gates it uses. First we expand it using the gate representations of \cref{Dirac_matrices}. One can then note that each of the $\gamma^i$ gates use the combination $ZX$ and thus one can simplify the full gate by placing an overall $ZX$ and compensating with an extra $Z$ gate to end up with only $X$ for $\gamma^0$ as $ZZ = \mathds{1}$. The expansion and gate reduction can be seen in \cref{A:multi-gamma}.\\

\begin{figure}
    \centering
    \includegraphics[width=0.45\linewidth]{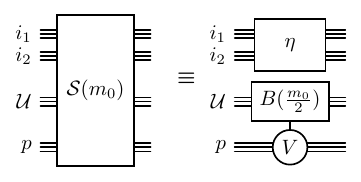}
    \caption{Circuit diagram for the single pole gate responsible for the first term in \cref{A:propagator}.}
    \label{fig:Sgate}
\end{figure}

\noindent\textbf{Propagator gate}\\
With the vertex gate designed we move to the development of the propagator gate. Recall that in $R_\xi$ gauge the propagator for the internal boson $V$ with four-momentum $k_\mu$ and mass $M_V$ is given by 
\begin{equation}
\begin{aligned}
    \Delta_{\mu\nu}(k)  = \frac{-i\eta_{\mu\nu}}{k^2 - M_V^2} + \frac{i(1 - \xi_V)k_\mu k_\nu}{(k^2 - M_V^2)(k^2 - \xi_VM_V^2)}
     \equiv m_0\eta_{\mu\nu} + m_1k_\mu k_\nu \label{A:propagator}
\end{aligned}
\end{equation}
where we have defined the single and double pole factors
\begin{equation}
\begin{aligned}
     m_0(s) \equiv \frac{-i}{k^2 - M_V^2} \andspace m_1(s) \equiv \frac{i(1 - \xi_V)}{(k^2 - M_V^2)(k^2 - \xi_VM_V^2)}. \label{A:poles}
\end{aligned}
\end{equation}
In the above we have the gauge choice parameter $\xi_V = 0, 1$ which refer to either Landau or Feynman gauge. For the construction of this gate we will keep it arbitrary and not make any gauge choice. Recall also that in the center-of-momentum (COM) frame the internal momentum squared gives $k^2 = s$ with $s$ being the Mandelstam variable. Of course this should be $\hat{s}$ for the partonic level scattering, however we make no distinction at this point and refer to it as $s$. In order to implement this propagator into our quantum state two simple gates responsible for each of these pole terms, the single $\Sing(m_0)$ and double $\D(m_1)$ pole gates are needed. Both of these gates will operate on the registers $\big\{i_1, i_2, \U, p\big\}$. The single pole gate $\Sing(m_0)$ has to apply the Minkowski metric tensor $\eta_{\mu\nu}$ and the pole factor $m_0$ and so we use the $\eta$ gate in \cref{fig:eta_circ} on the index registers and a controlled value setting gate $C_{|V\rangle}\big[B(m_0/2)\big]$ with the input rescaled by $1/2$ to compensate for the loss of this same factor in \cref{eta_action}. The value setting gate is controlled by the $p$ register so that different values of $m_0$ can be implemented depending on the type of particle. The circuit diagram for the single pole gate is shown in \cref{fig:Sgate}. 

\begin{figure}
    \centering
    \includegraphics[width=0.7\linewidth]{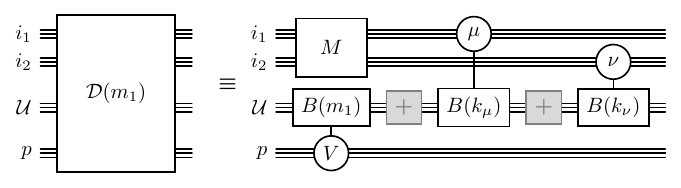}
    \caption{Circuit diagram for the double pole gate responsible for the second term in \cref{A:propagator}.}
    \label{fig:Dgate}
\end{figure}

In a similar, however, slightly more complicated manner, the double pole gate $\D(m_1)$ is designed to implement the double pole factor in \cref{A:propagator}. The scalar part of this term is just $m_1$ and so we use a value setting gate controlled by the particle register $C_{|V\rangle}\big[B(m_1)\big]$ for this. The tensorial part however is $k_\mu k_\nu$ and so we use value setting gates controlled by the index registers $C_{|\mu\rangle}\big[B(k_\mu)\big]C_{|\nu\rangle}\big[B(k_\nu)\big]$ to determine $\mu$ and $\nu$ as well as a metric signature gate to ensure the correct signs. We use the metric part of \cref{fig:eta_circ} and denote it by $M$ for this. The full double pole gate circuit diagram is shown in \cref{fig:Dgate}. 

\begin{figure}[h]
    \centering
    \includegraphics[width=0.75\linewidth]{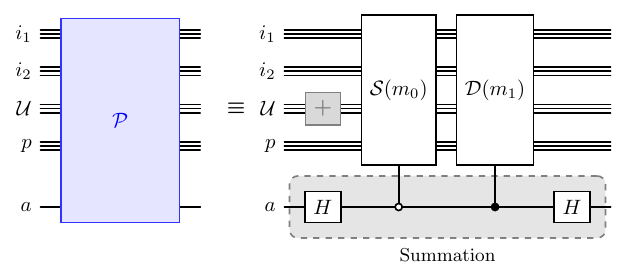}
    \caption{Circuit diagram for the decomposition of the propagator gate $\mathcal{P}$ which implements the propagator factor in \cref{A:propagator}.}
    \label{fig:Pgate}
\end{figure}

With these two gates ready, it is straightforward to construct the full propagator gate. As in the vertex gate construction we use an ancilla register that encodes $|0\rangle$ for the single term and $|1\rangle$ for the double term and then sums them together with an Hadamard gate. The full propagator gate is shown in \cref{fig:Pgate}. 

One final aspect that needs to be considered about this gate is the fact that the pole factors may have norms larger than one, of course rendering the value setting gates $B(\alpha)$ non-unitary. The fix for this is simple, just define a pole factor roof as the maximum of all the pole norms 
\begin{equation}
   \bar{m}(s_i) \equiv \text{max}_V\big\{|m_0(s_i)|, |m_1(s_i)|\big\}. \label{mroof}
\end{equation}
The subscript $V$ indicates that this roof should be for all the possible bosons. Once such a factor is found we rescale all the poles by $1/\bar{m}$ such that all the input values for the $B(\alpha)$ gates have norm equal to or less than one ensuring unitarity. The action of this is simply normalizing the input values, which also has the potential to improve the circuit's performance. This factor of $1/\bar{m}$ will only appear once in the final diagram and so it has to be compensated for in the final output rescaling as seen in \cref{compensation}. \\

\begin{figure}[h]
    \centering
    \includegraphics[width=0.6\linewidth]{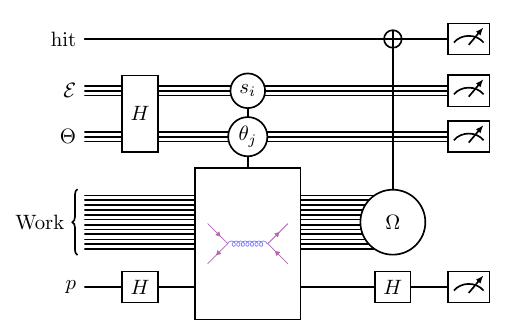}
    \caption{The simulated circuit with work vacuum projection to an ancilla hit register.}
    \label{fig:hitcircuit}
\end{figure}

\noindent \textbf{Reducing the number of simulated readout qubits}\\
As mentioned in the beginning of Section 3 we simulate a slightly altered circuit which is meant to reduce the number of readout qubits. We define work to be the set of registers $\{v_i, v_2, i_1, i_2, \U, \{a_i\}\}$ of the circuit in \cref{fig:PS-Drell-Yan_Circuit} and recall that the output state of interest is the vacuum state of these registers $\vac_{\text{work}}$. All the other states $\perp \vac_\text{work}$ are of non-interest for us and so we categorize a circuit output as hit if $\vac_\text{work}$ and miss if $\perp\vac_\text{work}$. In the simulation we introduce an ancilla qubit labeled as hit which is prepared in the vacuum. After the circuit in \cref{fig:PS-Drell-Yan_Circuit} we apply a Multi-CNOT on the work and hit register in such a way that the hit qubit flips if the $\vac_{\text{work}}$ state appears. The combined state in $\text{hit}\otimes \text{work}$ is then $\alpha |1\rangle_\text{hit}\vac_\text{work} + \beta|0\rangle_\text{hit}\otimes \big(\perp\vac_\text{work}\big)$. In this way we have projected the important information of the work registers onto one single qubit and we can skip the measurements of the latter. The full circuit that was simulated can be seen in \cref{fig:hitcircuit}.\\

\vspace*{0.5truecm}
\noindent\textbf{Scattering matrix elements}\\
The analytical expressions for the $\gamma$ and $Z$ diagrams in the $q\bar{q} \to \gamma/Z \to \mu^-\mu^+$ scattering are given by
\begin{equation}
  \mathcal{M}_\gamma = ie^2Q_qQ_\mu\frac{1}{s}\big[\bar{v}_2\gamma^\mu u_1\big]\big[\bar{u}_3\gamma_\mu v_4\big] \label{A:Mgamma}
\end{equation}
and
\begin{equation}
\begin{aligned}
    \mathcal{M}_Z & =  \frac{-i}{s - M_Z^2 + iM_Z\Gamma_Z}\bigg[\bar{v}_2\big(C_{\V, q}^{(Z)}\gamma^\mu + C_{\A, q}^{(Z)}\gamma^\mu \gamma_5 \big)u_1 \bigg]
  \bigg[\bar{u}_3\big(C_{\V, \mu}^{(Z)}\gamma_\mu + C_{\A, \mu}^{(Z)}\gamma_\mu \gamma_5 \big)v_4 \bigg]. \label{A:MZ}
\end{aligned}
\end{equation}

\vspace{3mm}
\section*{Appendix B: Circuit state derivation}    
\setcounter{equation}{0}
\renewcommand{\theequation}{B.\arabic{equation}}
We will below give a full derivation of the output quantum state of the circuit in \cref{fig:PS-Drell-Yan_Circuit}. The initial state of the circuit is with vacuum in all the registers:
\begin{equation}
    |\psi_0\rangle = \bigotimes_{r \in \{\E, \Theta, v_1, v_2, i_1, i_2, \U, a_1, a_2, a_3, p\}}|\Omega\rangle_r.
\end{equation}
Applying the first layer of Hadamards opens the superpositions of the $\E, \Theta, i_1, i_2$ and $p$ registers and gives
\begin{equation}
\begin{aligned}
    |\psi_1\rangle & = \sum_{\sqrt{s}_i}\frac{1}{\sqrt{2^{n_\E}}}|\sqrt{s}_i\rangle_\E \sum_{\theta_j}\frac{1}{\sqrt{2^{n_\Theta}}}|\theta_j\rangle_\Theta  \otimes \vac_{v_i} \vac_{v_2} \otimes \sum_{\mu}\frac{1}{2}|\mu\rangle_{i_1}\sum_{\nu}\frac{1}{2}|\nu\rangle_{i_2}\\
    & \hspace{5mm}\otimes  \vac_\U \bigotimes_{i = 1}^3\vac_{a_i} \otimes\sum_{V = \gamma, Z}\frac{1}{\sqrt{2}} |V\rangle_p.
\end{aligned}
\end{equation}
After passing through the first vertex the state has gained the vertex factor and spinors evaluated on the $v_1$ register which reads
\begin{equation}
    \begin{aligned}
        |\psi_2\rangle & = \frac{1}{\sqrt{2^{n_\E + n_\Theta + 6}}}\sum_{\{\sqrt{s}_i, \theta_j\}}\sum_{\mu, \nu}\sum_{V = \gamma, Z}\bar{U}_{p_2}\bigg(C_{\V, q}^{(V)}\gamma^\mu + C^{(V)}_{\A, q}\gamma^\mu\gamma_5\bigg)U_{p_1}\vac_{v_1} |\sqrt{s}_i\rangle_\E |\theta_j\rangle_\Theta |\mu\nu\rangle_{i_1 i_2}|V\rangle_p\\
        & \hspace{50mm}\otimes \vac_\U\vac_{v_2}\bigotimes_{i = 1}^3\vac_{a_i} \ortho{\vac_\U\vac_{a_1}}
    \end{aligned}
\end{equation}
where the parenthesis at the end indicates that all the other terms in the state are all orthogonal to $\vac_\U$ and $\vac_{a_1}$. Passing through the propagator the state gains the propagator factor controlled by the index registers and the energy register. 
\begin{equation}
    \begin{aligned}
         |\psi_3\rangle & = \frac{1}{\bar{m}\sqrt{2^{n_\E + n_\Theta + 7}}}\sum_{\{\sqrt{s}_i, \theta_j\}}\sum_{\mu, \nu}\sum_{V = \gamma, Z}\bar{U}_{p_2}\bigg(C_{\V, q}^{(V)}\gamma^\mu + C^{(V)}_{\A, q}\gamma^\mu\gamma_5\bigg)U_{p_1}\vac_{v_1} \\
         & \hspace{50mm}\otimes \bigg(m_0^{(V)}(s_i)\eta_{\mu\nu} + m_1^{(V)}(s_i)k_\mu k_\nu\bigg)|\sqrt{s}_i\rangle_\E |\theta_j\rangle_\Theta |\mu\nu\rangle_{i_1 i_2}|V\rangle_p\\
        & \hspace{50mm}\otimes \vac_\U\vac_{v_2}\bigotimes_{i = 1}^3\vac_{a_i} \ortho{\vac_\U\vac_{a_1}\vac_{a_2}}.
    \end{aligned}
\end{equation}
Note the scaling of $1/\bar{m}$ in the normalization as mentioned in section IIb. After the second vertex the quantum state has the form 
\begin{equation}
    \begin{aligned}
        |\psi_4\rangle & = \frac{1}{\bar{m}\sqrt{2^{n_\E + n_\Theta + 8}}}\sum_{\{\sqrt{s}_i, \theta_j\}}\sum_{\mu, \nu}\sum_{V = \gamma, Z}\bar{U}_{p_2}\bigg(C_{\V, q}^{(V)}\gamma^\mu + C^{(V)}_{\A, q}\gamma^\mu\gamma_5\bigg)U_{p_1}\vac_{v_1} \\
         & \hspace{50mm}\otimes \bigg(m_0^{(V)}(s_i)\eta_{\mu\nu} + m_1^{(V)}(s_i)k_\mu k_\nu\bigg)|\sqrt{s}_i\rangle_\E\\
         & \hspace{50mm} \otimes \bar{U}_{p_3}(\theta_j)\bigg(C_{\V, \mu}^{(V)}\gamma^\nu + C^{(V)}_{\A, \mu}\gamma^\nu\gamma_5\bigg)U_{p_4}(\theta_j)\vac_{v_2} |\theta_j\rangle_\Theta \\
        & \hspace{50mm}\otimes |\mu\nu\rangle_{i_1 i_2}|V\rangle_p\vac_\U\bigotimes_{i = 1}^3\vac_{a_i} \ortho{\vac_\U\bigotimes_{i = 1}^3\vac_{a_i}}.
    \end{aligned}
\end{equation}
The final step of the circuit is to close the index and particles registers with a final Hadamard transformation. The final state of the circuit is thus 
\begin{equation}
\begin{aligned}
    &  |\psi_\text{final}\rangle  = \frac{1}{\bar{m}\sqrt{2^{n_\E + n_\Theta + 12}}} \sum_{\{\sqrt{s}_i, \theta_j\}} 
    \Bigg\{
    \sum_{\mu, \nu}\sum_{V = \gamma, Z} \\
    & \bar{U}_{p_2}\bigg(C_{\V, q}^{(V)}\gamma^\mu + C^{(V)}_{\A, q}\gamma^\mu\gamma_5\bigg)U_{p_1}\vac_{v_1}\bigg(m_0^{(V)}(s_i)\eta_{\mu\nu} + m_1^{(V)}(s_i)k_\mu k_\nu\bigg) \bar{U}_{p_3}(\theta_j)\bigg(C_{\V, \mu}^{(V)}\gamma^\nu + C^{(V)}_{\A, \mu}\gamma^\nu\gamma_5\bigg)U_{p_4}(\theta_j)\vac_{v_2} \Bigg\}\\
         & \otimes |\sqrt{s}_i\rangle_\E |\theta_j\rangle_\Theta  
         \vac_p\vac_{\text{work}} \hspace{1mm }\ortho{\vac_p\vac_{\text{work}}}
\end{aligned}
\end{equation}
where the work register refer to the set $\{i_1, i_2, \U, \{a_i\}\}$. A final projection to a label state in $\E\otimes \Theta$ and the vacuum in the work gives the sum of the diagrams:

\begin{equation}
\begin{aligned}
    & \langle \sqrt{s}_i, \theta_j|\langle \Omega|_p\langle \Omega|_\text{work}|\psi_\text{final}\rangle  \sim  \M_\gamma(s_i, \theta_j) + \M_Z(s_i, \theta_j)
\end{aligned}
\end{equation}
up to the compensation factor $C(s_i)$ defined in \cref{compensation}.

\addcontentsline{toc}{section}{References}
\bibliographystyle{spphys_no_doi}
\bibliography{refs}

@article{Feuerstake:2025,
   title={Interference effects in resonant di-Higgs production at the LHC in the Higgs singlet extension},
   volume={2025},
   ISSN={1029-8479},
   url={http://dx.doi.org/10.1007/JHEP04(2025)094},
   DOI={10.1007/jhep04(2025)094},
   number={4},
   journal={Journal of High Energy Physics},
   publisher={Springer Science and Business Media LLC},
   author={Feuerstake, Finn and Fuchs, Elina and Robens, Tania and Winterbottom, Daniel},
   year={2025},
   month=Apr }

@article{Accomando:2013,
   title={Z′ at the LHC: interference and finite width effects in Drell-Yan},
   volume={2013},
   ISSN={1029-8479},
   url={http://dx.doi.org/10.1007/JHEP10(2013)153},
   DOI={10.1007/jhep10(2013)153},
   number={10},
   journal={Journal of High Energy Physics},
   publisher={Springer Science and Business Media LLC},
   author={Accomando, Elena and Becciolini, Diego and Belyaev, Alexander and Moretti, Stefano and Shepherd-Themistocleous, Claire},
   year={2013},
   month=Oct }

@article{mistlberger:2025,
    author = "Mistlberger, Bernhard and Suresh, Adi",
    title = "{RVV{\texttimes}V: interference contributions to inclusive Higgs boson and Drell-Yan production at N$^{4}$LO in QCD}",
    eprint = "2504.10574",
    archivePrefix = "arXiv",
    primaryClass = "hep-ph",
    reportNumber = "SLAC-PUB-250410",
    doi = "10.1007/JHEP11(2025)087",
    journal = "JHEP",
    volume = "11",
    pages = "087",
    year = "2025"
}

@article{Agliardi:2022,
   title={Quantum integration of elementary particle processes},
   volume={832},
   ISSN={0370-2693},
   url={http://dx.doi.org/10.1016/j.physletb.2022.137228},
   DOI={10.1016/j.physletb.2022.137228},
   journal={Physics Letters B},
   publisher={Elsevier BV},
   author={Agliardi, Gabriele and Grossi, Michele and Pellen, Mathieu and Prati, Enrico},
   year={2022},
   month=Sept, pages={137228} }

@article{Williams:2025,
   title={A general approach to quantum integration of cross sections in high-energy physics},
   volume={10},
   ISSN={2058-9565},
   url={http://dx.doi.org/10.1088/2058-9565/adf771},
   DOI={10.1088/2058-9565/adf771},
   number={4},
   journal={Quantum Science and Technology},
   publisher={IOP Publishing},
   author={Williams, Ifan and Pellen, Mathieu},
   year={2025},
   month=Aug, pages={045017} }

@article{varona:2024,
      title={Towards quantum computing Feynman diagrams in hybrid qubit-oscillator devices}, 
      author={S. Varona and S. Saner and O. Băzăvan and G. Araneda and G. Aarts and A. Bermudez},
      year={2024},
      eprint={2411.05092},
      archivePrefix={arXiv},
      primaryClass={quant-ph},
      url={https://arxiv.org/abs/2411.05092}, 
}

@article{Alwall:2014,
   title={The automated computation of tree-level and next-to-leading order differential cross sections, and their matching to parton shower simulations},
   volume={2014},
   ISSN={1029-8479},
   url={http://dx.doi.org/10.1007/JHEP07(2014)079},
   DOI={10.1007/jhep07(2014)079},
   number={7},
   journal={Journal of High Energy Physics},
   publisher={Springer Science and Business Media LLC},
   author={Alwall, J. and Frederix, R. and Frixione, S. and Hirschi, V. and Maltoni, F. and Mattelaer, O. and Shao, H.-S. and Stelzer, T. and Torrielli, P. and Zaro, M.},
   year={2014},
   month=July }

@article{javadiabhari:2024,
      title={Quantum computing with Qiskit}, 
      author={Ali Javadi-Abhari and Matthew Treinish and Kevin Krsulich and Christopher J. Wood and Jake Lishman and Julien Gacon and Simon Martiel and Paul D. Nation and Lev S. Bishop and Andrew W. Cross and Blake R. Johnson and Jay M. Gambetta},
      year={2024},
      eprint={2405.08810},
      archivePrefix={arXiv},
      primaryClass={quant-ph},
      url={https://arxiv.org/abs/2405.08810}, 
}

@article{Klco:2021lap,
    author = "Klco, Natalie and Roggero, Alessandro and Savage, Martin J.",
    title = "{Standard model physics and the digital quantum revolution: thoughts about the interface}",
    eprint = "2107.04769",
    archivePrefix = "arXiv",
    primaryClass = "quant-ph",
    reportNumber = "IQuS@UW-21-007",
    doi = "10.1088/1361-6633/ac58a4",
    journal = "Rept. Prog. Phys.",
    volume = "85",
    number = "6",
    pages = "064301",
    year = "2022"
}

@article{Funcke:2023jbq,
    author = {Funcke, Lena and Hartung, Tobias and Jansen, Karl and K{\"u}hn, Stefan},
    title = "{Review on Quantum Computing for Lattice Field Theory}",
    eprint = "2302.00467",
    archivePrefix = "arXiv",
    primaryClass = "hep-lat",
    reportNumber = "MIT-CTP/5482",
    doi = "10.22323/1.430.0228",
    journal = "PoS",
    volume = "LATTICE2022",
    pages = "228",
    year = "2023"
}

@article{Yamamoto:2022jnn,
    author = "Yamamoto, Arata and Doi, Takumi",
    title = "{Toward Nuclear Physics from Lattice QCD on Quantum Computers}",
    eprint = "2211.14550",
    archivePrefix = "arXiv",
    primaryClass = "hep-lat",
    reportNumber = "RIKEN-iTHEMS-Report-22",
    doi = "10.1093/ptep/ptae019",
    journal = "PTEP",
    volume = "2024",
    number = "3",
    pages = "033D02",
    year = "2024"
}

@article{Meurice:2020pxc,
    author = "Meurice, Yannick and Sakai, Ryo and Unmuth-Yockey, Judah",
    title = "{Tensor lattice field theory for renormalization and quantum computing}",
    eprint = "2010.06539",
    archivePrefix = "arXiv",
    primaryClass = "hep-lat",
    reportNumber = "FERMILAB-PUB-20-580-QIS-T",
    doi = "10.1103/RevModPhys.94.025005",
    journal = "Rev. Mod. Phys.",
    volume = "94",
    number = "2",
    pages = "025005",
    year = "2022"
}

@article{deLejarza:2024pgk,
    author = "de Lejarza, Jorge J. Mart{\'\i}nez and Cieri, Leandro and Grossi, Michele and Vallecorsa, Sofia and Rodrigo, Germ{\'a}n",
    title = "{Loop Feynman integration on a quantum computer}",
    eprint = "2401.03023",
    archivePrefix = "arXiv",
    primaryClass = "hep-ph",
    doi = "10.1103/PhysRevD.110.074031",
    journal = "Phys. Rev. D",
    volume = "110",
    number = "7",
    pages = "074031",
    year = "2024"
}

@article{Ramirez-Uribe:2021ubp,
    author = "Ram{\'\i}rez-Uribe, Selomit and Renter{\'\i}a-Olivo, Andr{\'e}s E. and Rodrigo, Germ{\'a}n and Sborlini, German F. R. and Vale Silva, Luiz",
    title = "{Quantum algorithm for Feynman loop integrals}",
    eprint = "2105.08703",
    archivePrefix = "arXiv",
    primaryClass = "hep-ph",
    reportNumber = "IFIC/21-15, DESY 21-067",
    doi = "10.1007/JHEP05(2022)100",
    journal = "JHEP",
    volume = "05",
    pages = "100",
    year = "2022"
}

@article{Clemente:2022nll,
    author = "Clemente, Giuseppe and Crippa, Arianna and Jansen, Karl and Ram{\'\i}rez-Uribe, Selomit and Renter{\'\i}a-Olivo, Andr{\'e}s E. and Rodrigo, Germ{\'a}n and Sborlini, German F. R. and Vale Silva, Luiz",
    title = "{Variational quantum eigensolver for causal loop Feynman diagrams and directed acyclic graphs}",
    eprint = "2210.13240",
    archivePrefix = "arXiv",
    primaryClass = "hep-ph",
    reportNumber = "IFIC/22-28",
    doi = "10.1103/PhysRevD.108.096035",
    journal = "Phys. Rev. D",
    volume = "108",
    number = "9",
    pages = "096035",
    year = "2023"
}

@article{Bauer:2021gup,
    author = "Bauer, Christian W. and Freytsis, Marat and Nachman, Benjamin",
    title = "{Simulating Collider Physics on Quantum Computers Using Effective Field Theories}",
    eprint = "2102.05044",
    archivePrefix = "arXiv",
    primaryClass = "hep-ph",
    doi = "10.1103/PhysRevLett.127.212001",
    journal = "Phys. Rev. Lett.",
    volume = "127",
    number = "21",
    pages = "212001",
    year = "2021"
}

@article{Perez-Salinas:2020nem,
    author = "P{\'e}rez-Salinas, Adri{\'a}n and Cruz-Martinez, Juan and Alhajri, Abdulla A. and Carrazza, Stefano",
    title = "{Determining the proton content with a quantum computer}",
    eprint = "2011.13934",
    archivePrefix = "arXiv",
    primaryClass = "hep-ph",
    reportNumber = "TIF-UNIMI-2020-30",
    doi = "10.1103/PhysRevD.103.034027",
    journal = "Phys. Rev. D",
    volume = "103",
    number = "3",
    pages = "034027",
    year = "2021"
}

@article{Li:2024zsw,
    author = "Li, Tianyin and Xing, Hongxi",
    title = "{Partonic collinear structure by quantum computing}",
    doi = "10.22323/1.469.0029",
    journal = "PoS",
    volume = "DIS2024",
    pages = "029",
    year = "2025"
}

@article{Bepari:2020xqi,
    author = "Bepari, Khadeejah and Malik, Sarah and Spannowsky, Michael and Williams, Simon",
    title = "{Towards a quantum computing algorithm for helicity amplitudes and parton showers}",
    eprint = "2010.00046",
    archivePrefix = "arXiv",
    primaryClass = "hep-ph",
    reportNumber = "IPPP/20/41",
    doi = "10.1103/PhysRevD.103.076020",
    journal = "Phys. Rev. D",
    volume = "103",
    number = "7",
    pages = "076020",
    year = "2021"
}

@article{Bepari:2021kwv,
    author = "Bepari, Khadeejah and Malik, Sarah and Spannowsky, Michael and Williams, Simon",
    title = "{Quantum walk approach to simulating parton showers}",
    eprint = "2109.13975",
    archivePrefix = "arXiv",
    primaryClass = "hep-ph",
    doi = "10.1103/PhysRevD.106.056002",
    journal = "Phys. Rev. D",
    volume = "106",
    number = "5",
    pages = "056002",
    year = "2022"
}

@article{Agliardi:2022ghn,
    author = "Agliardi, Gabriele and Grossi, Michele and Pellen, Mathieu and Prati, Enrico",
    title = "{Quantum integration of elementary particle processes}",
    eprint = "2201.01547",
    archivePrefix = "arXiv",
    primaryClass = "hep-ph",
    reportNumber = "FR-PHENO-2022-01",
    doi = "10.1016/j.physletb.2022.137228",
    journal = "Phys. Lett. B",
    volume = "832",
    pages = "137228",
    year = "2022"
}

@article{Williams:2025hza,
    author = "Williams, Ifan and Pellen, Mathieu",
    title = "{A general approach to quantum integration of cross sections in high-energy physics}",
    eprint = "2502.14647",
    archivePrefix = "arXiv",
    primaryClass = "quant-ph",
    reportNumber = "FR-PHENO-2025-02",
    doi = "10.1088/2058-9565/adf771",
    journal = "Quantum Sci. Technol.",
    volume = "10",
    number = "4",
    pages = "045017",
    year = "2025"
}

@article{Gustafson:2022dsq,
    author = {Gustafson, G{\"o}sta and Prestel, Stefan and Spannowsky, Michael and Williams, Simon},
    title = "{Collider events on a quantum computer}",
    eprint = "2207.10694",
    archivePrefix = "arXiv",
    primaryClass = "hep-ph",
    doi = "10.1007/JHEP11(2022)035",
    journal = "JHEP",
    volume = "11",
    pages = "035",
    year = "2022"
}

@article{Bravo-Prieto:2021ehz,
    author = "Bravo-Prieto, Carlos and Baglio, Julien and C{\`e}, Marco and Francis, Anthony and Grabowska, Dorota M. and Carrazza, Stefano",
    title = "{Style-based quantum generative adversarial networks for Monte Carlo events}",
    eprint = "2110.06933",
    archivePrefix = "arXiv",
    primaryClass = "quant-ph",
    reportNumber = "CERN-TH-2021-139, TIF-UNIMI-2021-14",
    doi = "10.22331/q-2022-08-17-777",
    journal = "Quantum",
    volume = "6",
    pages = "777",
    year = "2022"
}

@article{Kiss:2022pjw,
    author = "Kiss, Oriel and Grossi, Michele and Kajomovitz, Enrique and Vallecorsa, Sofia",
    title = "{Conditional Born machine for Monte Carlo event generation}",
    eprint = "2205.07674",
    archivePrefix = "arXiv",
    primaryClass = "quant-ph",
    doi = "10.1103/PhysRevA.106.022612",
    journal = "Phys. Rev. A",
    volume = "106",
    number = "2",
    pages = "022612",
    year = "2022"
}

@article{Bauer:2022hpo,
    author = "Bauer, Christian W. and others",
    title = "{Quantum Simulation for High-Energy Physics}",
    eprint = "2204.03381",
    archivePrefix = "arXiv",
    primaryClass = "quant-ph",
    reportNumber = "UMD-PP-022-04, LA-UR-22-22100, RIKEN-iTHEMS-Report-22, RIKEN-iTHEMS-Report-22,
  FERMILAB-PUB-22-249-SQMS-T, IQuS@UW-21-027, MITRE-21-03848-2, FERMILAB-PUB-22-249-SQMS-T",
    doi = "10.1103/PRXQuantum.4.027001",
    journal = "PRX Quantum",
    volume = "4",
    number = "2",
    pages = "027001",
    year = "2023"
}

@article{DiMeglio:2023nsa,
    author = "Di Meglio, Alberto and others",
    title = "{Quantum Computing for High-Energy Physics: State of the Art and Challenges}",
    eprint = "2307.03236",
    archivePrefix = "arXiv",
    primaryClass = "quant-ph",
    reportNumber = "FERMILAB-PUB-23-468-ETD",
    doi = "10.1103/PRXQuantum.5.037001",
    journal = "PRX Quantum",
    volume = "5",
    number = "3",
    pages = "037001",
    year = "2024"
}

@article{Chawdhry:2023jks,
    author = "Chawdhry, Herschel A. and Pellen, Mathieu",
    title = "{Quantum simulation of colour in perturbative quantum chromodynamics}",
    eprint = "2303.04818",
    archivePrefix = "arXiv",
    primaryClass = "hep-ph",
    reportNumber = "FR-PHENO-2023-02, OUTP-23-02P",
    doi = "10.21468/SciPostPhys.15.5.205",
    journal = "SciPost Phys.",
    volume = "15",
    number = "5",
    pages = "205",
    year = "2023"
}

@article{Bashore:2025uwb,
    author = "Bashore, Erik and Moretti, Stefano and Vitos, Timea",
    title = "{A quantum algorithm for the n-gluon MHV scattering amplitude}",
    eprint = "2507.14252",
    archivePrefix = "arXiv",
    primaryClass = "hep-ph",
    doi = "10.21468/SciPostPhys.20.4.114",
    journal = "SciPost Phys.",
    volume = "20",
    number = "4",
    pages = "114",
    year = "2026"
}

\end{document}